\documentclass[a4paper]{article}
\usepackage[english]{babel}
\usepackage[utf8x]{inputenc}
\usepackage[T1]{fontenc}
\usepackage{afterpage}
\usepackage{float}
\usepackage{setspace}

\usepackage[a4paper,top=3cm,bottom=2cm,left=3cm,right=3cm,marginparwidth=1.75cm]{geometry}

\usepackage{amsmath}
\usepackage{graphicx}
\usepackage[colorinlistoftodos]{todonotes}
\usepackage[colorlinks=true, allcolors=blue]{hyperref}
\usepackage{csvsimple}
\usepackage{pdflscape}
\usepackage{mathptmx}
\begin{document}

\title{Textual analysis of artificial intelligence manuscripts reveals features associated with peer review outcome.}

\author{Philippe Vincent-Lamarre \textsuperscript{1,2,*}         \and
        Vincent Larivière \textsuperscript{1}    
}

\maketitle

\begin{enumerate}
\item[1] École de bibliothéconomie et des sciences de l'information, Université de Montréal, Montréal, Canada
\item[2] École de psychologie, Université d'Ottawa, Ottawa, Canada
\item[*] Corresponding author
\end{enumerate}

%
%

\section*{Abstract} 

We analysed a dataset of scientific manuscripts that were submitted to various conferences in artificial intelligence. We performed a combination of semantic, lexical and psycholinguistic analyses of the full text of the manuscripts and compared them with the outcome of the peer review process. We found that accepted manuscripts scored lower than rejected manuscripts on two indicators of readability, and that they also used more scientific and artificial intelligence jargon. We also found that accepted manuscripts were written with words that are less frequent, that are acquired at an older age, and that are more abstract than rejected manuscripts. The analysis of references included in the manuscripts revealed that the subset of accepted submissions were more likely to cite the same publications. This finding was echoed by pairwise comparisons of the word content of the manuscripts (i.e. an indicator or semantic similarity), which were more similar in the subset of accepted manuscripts. Finally, we predicted the peer review outcome of manuscripts with their word content, with words related to machine learning and neural networks positively related with acceptance, whereas words related to logic, symbolic processing and knowledge-based systems negatively related with acceptance.

\section{Introduction}

Peer review is a fundamental component of the scientific enterprise and acts as one of the main source of quality control of the scientific literature \cite{ziman_real_2002}. The primary form of peer review occurs before publication \cite{wakeling_no_2019} and it is often considered as a stamp of approval from the scientific community \cite{mulligan_is_2005,mayden_peer_2012}. Peer-reviewed publications have a considerable weight in the attribution of research and academic resources \cite{tregellas_predicting_2018,mckiernan_use_2019,moher_assessing_2018}.  

One of the main concern about peer review is its lack of reliability \cite{bailar_reliability_1991,lee_kuhnian_2012,cicchetti_reliability_1991}. Most studies on the topic find that agreement between reviewers is barely greater than chance \cite{bornmann_reliability-generalization_2010,price_nips_2014,forscher_how_2019}, which highlights the considerable amount of subjectivity involved in the process. This leaves room for a lot of potential source of bias, which have been reported in several studies \cite{lee_bias_2013,de_silva_preserving_2017,murray_gender_2018}. A potential silver lining is that it appears that the process has some validity. For instance, articles accepted at a general medicine journal \cite{jackson_validity_2011} and journals in the domain of ecology \cite{paine_effectiveness_2018} were more cited than the rejected articles published elsewhere, and the process appears to improve the quality of manuscripts, although marginally \cite{goodman_manuscript_1994,pierie_readers_1996,calcagno_flows_2012}. It is therefore surprising that a process that has little empirical support of its effectiveness, but a lot of evidence of its downsides \cite{smith_classical_2010} has so much importance. 

The vast majority of studies on peer review have focused on the relationship between the socio-demographical attributes of the actors involved in the process and its outcome \cite{sabaj_meruane_what_2016}. Comparatively, little research has focused on the association between the content of the manuscripts and the peer review process. This isn't surprising given that there is little publicly available datasets of manuscripts annotated as rejected or accepted, and whenever they are made available to researchers it is usually through smaller samples designed to answer specific questions. Another factor contributing to this gap in the litterature is that it is more time consuming to analyse textual data (either the referee's report or the reviewed manuscript) than papers' metadata. However, the increasing popularity of open access \cite{piwowar_state_2018,sutton_popularity_2017} allows for a greater access to the full text of scientific manuscripts.

By scraping the content of arXiv, one of those repositories,  \cite{kang_dataset_2018} developed a new method to identify manuscripts that were accepted at conferences after the peer review process based on submissions around the time of major NLP, machine learning (ML) and artificial intelligence (AI) conferences. These pre-prints were then matched with manuscripts that were published at the target venues as a way to determine whether they were accepted or "probably-rejected". In addition, the manuscripts and peer-review outcomes were collected from conferences that agreed to share their data.
 \cite{kang_dataset_2018} were able to achieve decent accuracy at predicting the acceptance of the manuscripts in their dataset. Other groups were able to obtain good performance at predicting paper acceptance with different machine learning models based on the text of the manuscripts \cite{jen_predicting_2018}, sentiment analysis of referee's reports \cite{ghosal_sentiment_2019}, or the evaluation score given by the reviewers \cite{qiao_modularized_2018}.

In this manuscript, we take advantage of the full text access to those manuscripts and explore linguistic and semantic features that correlates with the peer review outcome. Such features are relevant to two types of biases that could be involved in the peer review process: language and content bias. In the language bias, authors who aren't native English speakers could receive more negative evaluations due to the linguistic level of their manuscripts \cite{ross_effect_2006,tregenza_gender_2002,herrera_language_1999}. Our understanding of the extent to which such a bias could play a role in research evaluation is still limited, which is worrying given the increasingly globalized scientific system that relies on one language: English \cite{lariviere_introduction:_2019}. In terms of content bias, innovative and unorthodox methods are less likely to be judged favourably \cite{lee_bias_2013}. This type of bias is also quite likely to play a role in fields that are dominated by a few mainstream approaches such as AI \cite{hao_we_2019}. Conservatism in this field could impede the emergence of breakthrough or novel techniques that don't fit with the current trends.

In this manuscript, we address both types of biases by comparing the textual data (title, abstract and introduction) of the manuscripts. We first used two readability metrics (the Flesch Reading Ease (FRE) and the New Dale-Chall Readability (NDC) Formula), as well as some indicators of scientific jargon content, and found that manuscripts that were less readable and used more jargon were more likely to get accepted. Accepted and rejected manuscripts were compared on their psycholinguistic and lexical attributes and we found that accepted manuscripts used words that were more abstract, less frequent and acquired at a later age compared to rejected manuscripts. We then compared manuscripts on their word content and their referencing patterns through bibliographic coupling, and found that the subset of accepted manuscripts were semantically closer than rejected manuscripts. Finally, we used the word content of the manuscripts to predict their acceptance, and found that specific topics were associated with greater odds of acceptance. 

\section{Methods}

\subsection{Manuscript data}
We used the publicly available PeerRead dataset \cite{kang_dataset_2018} to analyse the semantic and lexical differences between accepted and rejected submissions to some natural language processing, artificial intelligence and machine learning conferences. We therefore used content from six platforms archived in the PeerRead dataset: three arXiv sub-repositories tagged by subject including submissions from 2007 to 2017 (AI: artificial intelligence, CL: computation and language, LG: machine learning), as well as submissions to three other venues: (ACL 2017: Association for Computational Linguistics, CoNLL 2016: Conference on Computational Natural Language Learning, ICLR 2017: International Conference on Learning Representations). This resulted in a dataset with 12,364 submissions. Although the submissions to ACL 2017 and CoNLL 2016 had an acceptance rate in \cite{kang_dataset_2018}, the information for each submission was not available in the dataset at the time of the analysis.

We limited our analysis to the title, abstract and introduction of the manuscripts, because the methods and results contained formulas, mathematical equations and variables, which made them unsuitable for textual analysis.

\begin{table}[htbp!]
\centering
\begin{tabular}{l|r|r}
Platform & \# Papers & \# Accepted \\\hline
ICLR 2017 & 427 & 172\\
ACL 2017 & 137 & -\\
CoNLL 2016 & 22 & -\\ 
arXiv:ai & 4092 & 418\\
arXiv:cl & 2638 & 646\\
arXiv:lg & 5048 & 1827\\
\end{tabular}
\caption{\label{tab:papers} Number of papers per platform.}
\end{table}

\subsection{Semantic similarity}
The textual data of each article, including the title, abstract, introduction were cleaned by making all words lowercase, eliminating punctuation, single character words and common stopwords. For all analyses except for the readability, scientific jargon and psycholinguistic matching, the stem of the word was extracted using the porter algorithm \cite{porter_algorithm_1980}. We used the Term Frequency Inverse Document Frequency (tf-idf) algorithm to create vectorial representations based on the field of interest (title, abstract or introduction). We then used those vectors to compute the cosine similarity between the pairs of documents.

\subsection{Reference matching}
In order to obtain manuscript's bibliographic coupling, we developed a reference matching algorithm because their format was not standardized across manuscripts. We used four conditions to group references together:
1- They were published the same year
2- They had the same number of authors
3- They had a similarity score above 0.7 (empirically determined after manual inspection of matching results) with a fuzzy matching procedure (\textit{Token Set Ratio} function from the FuzzyWuzzy python library, https://github.com/seatgeek/fuzzywuzzy) on the author's names and 4- the article's title.

\subsection{Psycholinguistic and readability variables}
For the word frequency estimation, we used the SUBTLEX\textsubscript{US} corpus \cite{brysbaert_moving_2009} from which we used the logarithm of the estimated word frequency + 1. For the concreteness, we used the \cite{brysbaert_concreteness_2014} dataset providing concreteness rating for 40,000 commonly known English words. For the age of acquisition, we used the \cite{kuperman_age--acquisition_2012} age of acquisition ratings for 30,000 English words.

We used the readability functions as implemented in \cite{plaven-sigray_readability_2017}. We used the Flech Reading Ease (FRE; \cite{flesch_new_1948,kincaid_derivation_1975}) and the New Dale-Chall Readability Formula (NDC; \cite{chall_readability_1995}). The FRE is calculated based on the number of syllables per words and the number of words per sentence. The NDC is based on the number of words per sentence and the proportion of difficult words that are not part of a list of "common words". We also included two sources of jargon developed by \cite{plaven-sigray_readability_2017}. The first one are science-specific common words, which are words used by scientist which are not in the NDC's list of common words. The other is the general science jargon, which are words frequently used in science, but aren't specific to science (see \cite{plaven-sigray_readability_2017} for methods). Finally, we complied a list of AI jargon from three online glossaries ($https://developers.google.com/machine-learning/glossary/$ , $http://www.wildml.com/deep-learning-glossary/$ and $https://en.wikipedia.org/wiki/Glossary\_of\_artificial\_intelligence$). 

\subsection{Data analysis}
Because of the exploratory nature of the study and the large size of the datasets, null hypothesis significance testing has many shortcomings \cite{szucs_when_2017}. In some cases, we performed statistical analysis of the results and reported the p-value, but those results should be interpreted carefully. Our analyses rely on the effect size, as well as the cross-validated effects on the independent subsets of the PeerRead dataset (manuscripts from different venues and online repositories). All error bars represent the standard error of the mean.

\subsection{Identification of geographic location}
We searched through the email addresses of the authors to identify research within the United-States (US) and outside the US. We considered a manuscript as US based if at least one author had an email address that ended with ".edu". 

\subsection{Code and data availability}
The custom python scripts and the data used to generate the results of this manuscript can be found at \url{https://github.com/lamvin/PeerReviewAI.git}.

\section{Results}

\subsection{Readability}

The readability of scientific articles has been steadily declining in the last century \cite{plaven-sigray_readability_2017}. One possible explanation for this is that writing more complex sentences and using more scientific jargon increase the likelihood that a manuscript will get accepted at peer review. To investigate this hypothesis, we used two measures of readability on our data: the Flesch Reading Ease (FRE) and the New Dale-Chall Readability Formula (NDC). FRE scores decrease as a function of a ratio of the number of syllables per word and the number of word per sentence. NDC scores increase as a function of the number of words in each sentence and as the proportion of difficult words increase (words that are not present in the NDC list of common words). We also included the proportion of words from a science-specific common words and general science jargon list (constructed by \cite{plaven-sigray_readability_2017}). In order to control for potential demographic confounds, we divided our dataset in the two categories: manuscripts within and outside the United States (US).  

\begin{figure}
\centering
\includegraphics[width=1\textwidth]{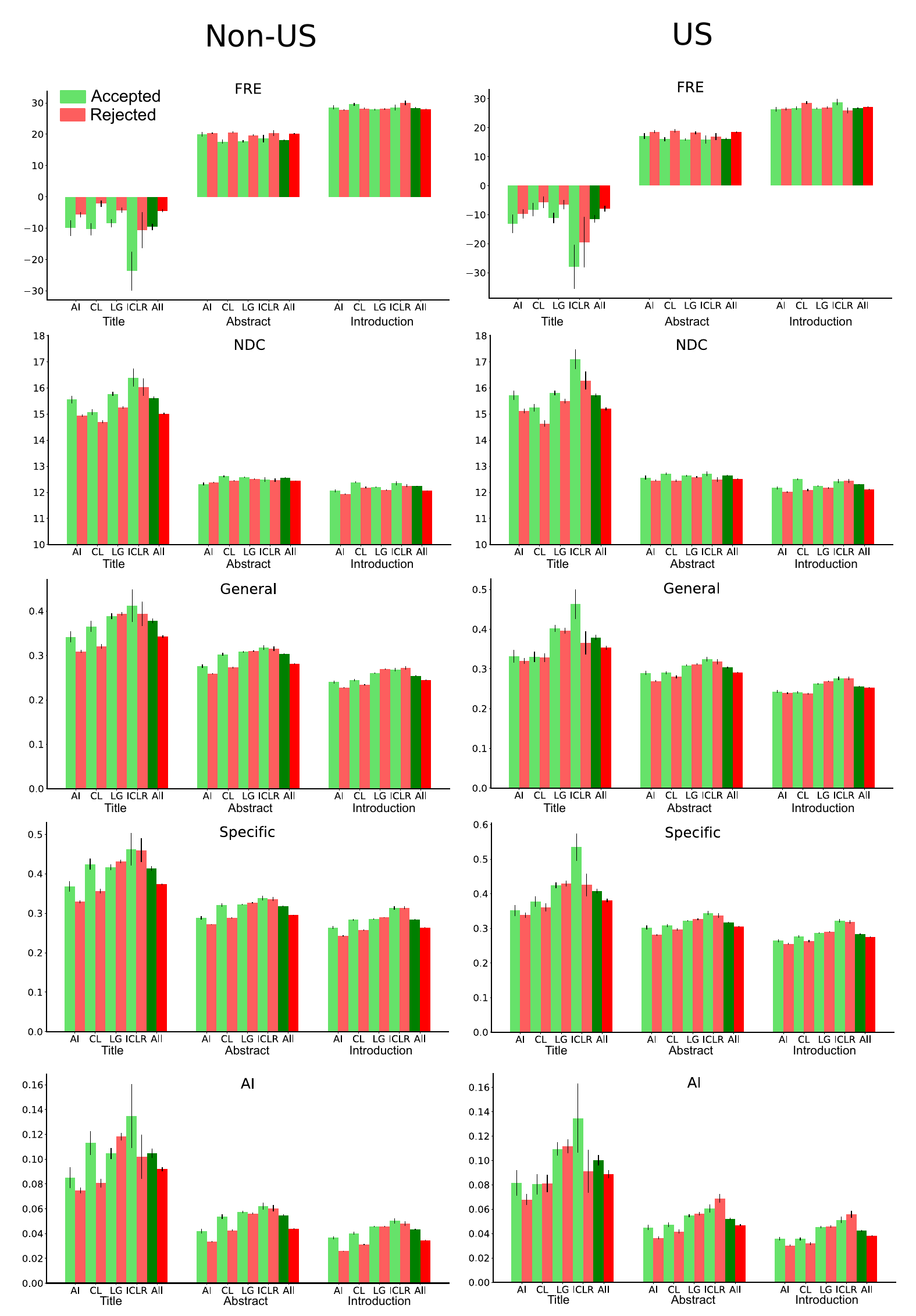}
\caption{\label{fig:Readability} Average readability and jargon proportion of US and non-US manuscripts.}
\end{figure}

We found that both indicators of readability were correlated with the peer review outcome. FRE (higher score = more readable) was lower for accepted manuscripts, while NDC (higher score = less readable) was higher for accepted manuscripts (Fig. \ref{fig:Readability}). This was the case for both US and non-US manuscripts (with no sizeable differences), for every section of the manuscript (title, abstract and introduction) and the effect was replicated within most platforms, except for the introduction with the FRE indicator. 

As \cite{plaven-sigray_readability_2017} reported that the proportion of scientific jargon has increased over the last century, we also wondered if the peer review process would reflect this effect. Using a list of general and specific scientific jargon, we found that the manuscripts containing the higher ratios of jargon were associated with higher acceptance rates. However, our list of science jargon was biased towards content from the life sciences. To confirm the relevance of these results to our dataset, we generated an AI jargon list (see Methods). Using this new list, we found a robust effect across platforms and document section, where a larger proportion of AI jargon predicted greater odds of acceptance for the manuscripts.

The replication of our results independently for US and non-US based manuscripts suggest that the effect is not driven by geographic locations. Statistical analysis are summarised in Tables \ref{Read_US} and \ref{Read_non_US}.

\subsection{Lexical correlates of peer review outcome}
We then investigated the differences between accepted and rejected submissions based on lexical and psycholinguistic attributes. Given that we haven't found consistent differences between US and non-US based submissions, we pooled all manuscripts together for the rest of the analysis. We used the number of tokens (total number of words in a document) as well as two measures of lexical diversity: the number of types (unique words in a document) and the ratio between the types and token (Type-Token Ratio, TTR). We also used three psycholinguistic variables: the age of acquisition (AOA), concreteness and frequency (on a logarithmic scale). We computed the average values of those psycholinguistic variables on all types and all tokens. 

\begin{figure}[H]
\centering
\includegraphics[width=1.1\textwidth]{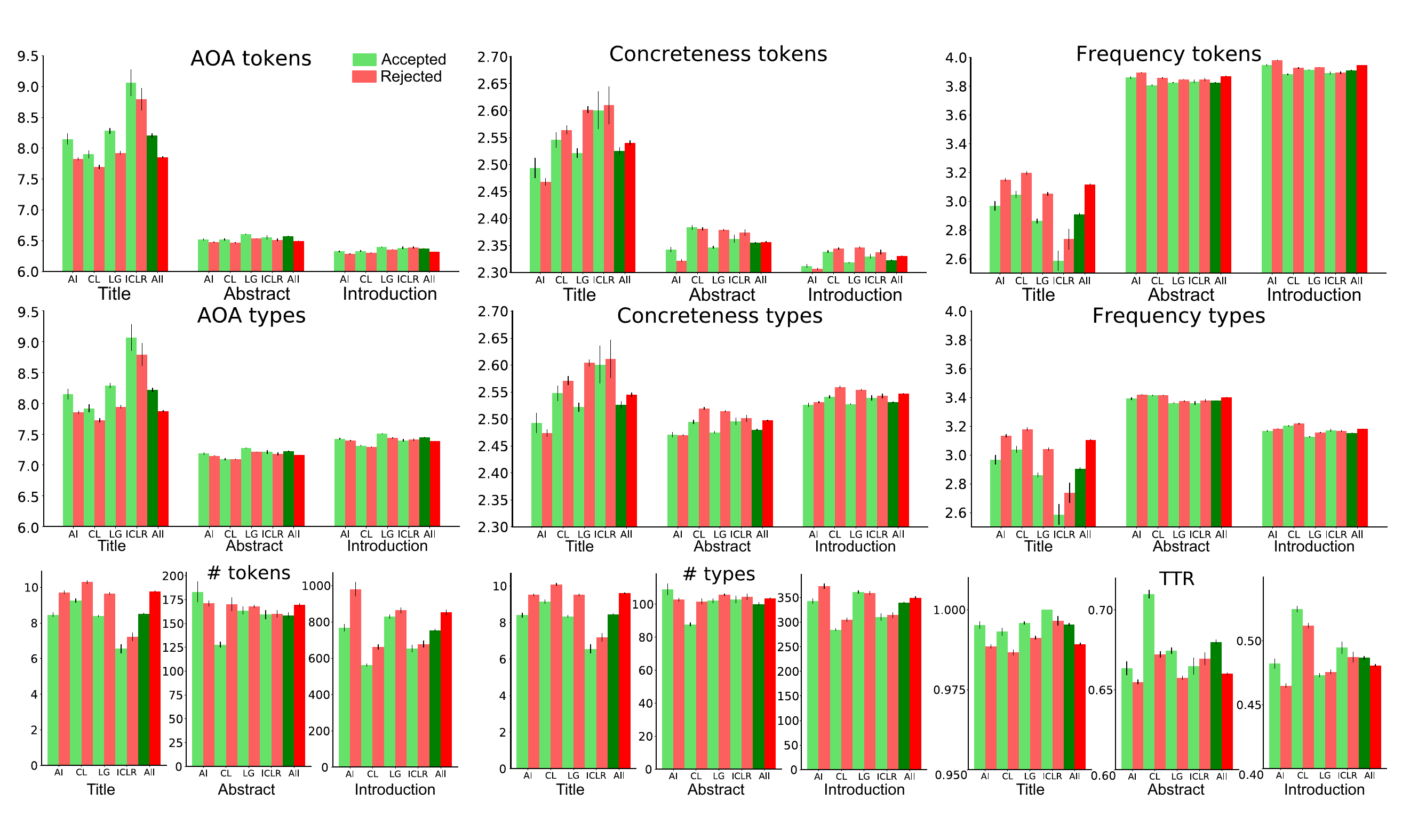}
\caption{\label{fig:PR_psych} Psycholinguistic and lexical scores of manuscripts.}
\end{figure}

We found consistent effects between the psycholinguistic variables and across platforms and section, with few exceptions. Words used in accepted manuscripts were less frequent, acquired later in life and more abstract than in rejected manuscripts on average (Fig. \ref{fig:PR_psych}). The effects were consistent across all platforms except ICLR (which is smaller than the other ones). 

Interestingly, we found that shorter titles, abstract and introduction were all associated with higher acceptance rates. Unsurprisingly, this translated in a bias towards manuscript with lower total types (for every section) for manuscript acceptance. However, when taking the ratio between the two (TTR, an indicator of lexical richness), we found that this variable was positively associated with manuscript acceptance. Results from the statistical analysis are summarised in Table \ref{Linguistic}.

\subsection{High-level semantic correlates of peer review outcome}
\subsubsection{Bibliographic coupling and semantic similarity}
We then looked at how similar the accepted manuscripts were compared to the rejected ones based on their semantic content. First we looked at the similarity of their title, abstract or introduction based on a tf-idf representation of their word content. Secondly, we looked at their degree of bibliographic coupling. We only compared pairs of manuscripts that shared at least one common reference for the next analysis (see table \ref{tab:references}). 

\begin{table}[htbp!]
\centering
\begin{tabular}{l|r}
\# of common cited references & Frequency \\\hline
1 & 1,619,173 \\
2 & 379,161 \\
3 & 124,574 \\ 
4 & 47,254 \\
5-9 & 36,753 \\
10-19 & 2,527 \\
20-29 & 230 \\ 
30-39 & 68 \\
40-49 & 31 \\ 
50-59 & 11 \\
60-80 & 4
\end{tabular}
\caption{\label{tab:references} Citation intersection for all papers.}
\end{table}

\begin{figure}[H]
\centering
\includegraphics[width=1\textwidth]{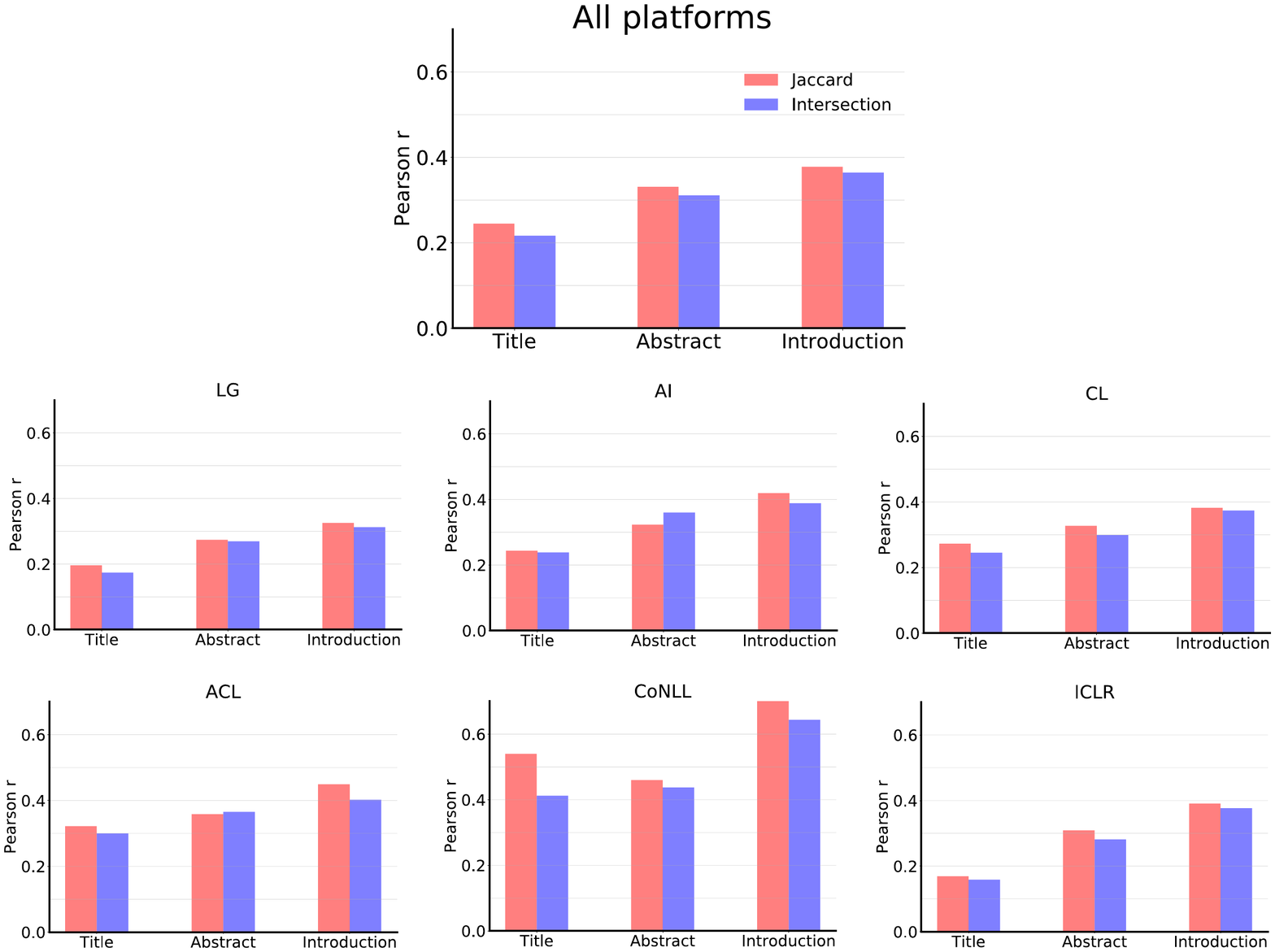}
\caption{\label{fig:BC_text_sim} Correlation between semantic similarity and bibliographic coupling.}
\end{figure}

As the two approaches quantify the content similarity of the documents, we wanted to verify whether those two metrics measured different aspects of the document content. It was previously reported that there is a moderate correlation between the two measures in the field of economics \cite{sainte-marie_you_2018}. We correlated the semantic distance with the bibliographic coupling of the document submitted to each platform. We used a semantic distance metric based on the cosine similarity between the tf-idf representation of each document, as well as both the citation intersection (\# common references) and the Jaccard similarity coefficient (\#references in common/ \# references in total) as a measure of bibliographic coupling. We found a moderate correlation (Pearson \textit{r} > 0.20 and < 0.40) between both measures of bibliographic coupling and semantic distance when pooling all platforms together depending on what section of the manuscript were compared (Fig. \ref{fig:BC_text_sim}). This suggests that those two measures aren't redundant features of semantic content, and that they might capture different aspects of it. This also validates our algorithm for citation disambiguation as comparable correlations between the bibliographic coupling and textual similarity were reported in \cite{sainte-marie_you_2018}.

\subsubsection{Bibliographic coupling and peer review outcome}
We then looked at how accepted and rejected manuscripts differed based on the characteristics of their cited references (bibliographic coupling). We compared all pairs of manuscripts on the two indicators of bibliographic coupling (intersection and Jaccard index). Each pair of manuscripts was categorized as one of the following: "accepted": the two submissions were accepted, "rejected": the two submissions were rejected, and "mixed", one document was rejected and the other was accepted.

\begin{figure}[H]
\centering
\includegraphics[width=0.65\textwidth]{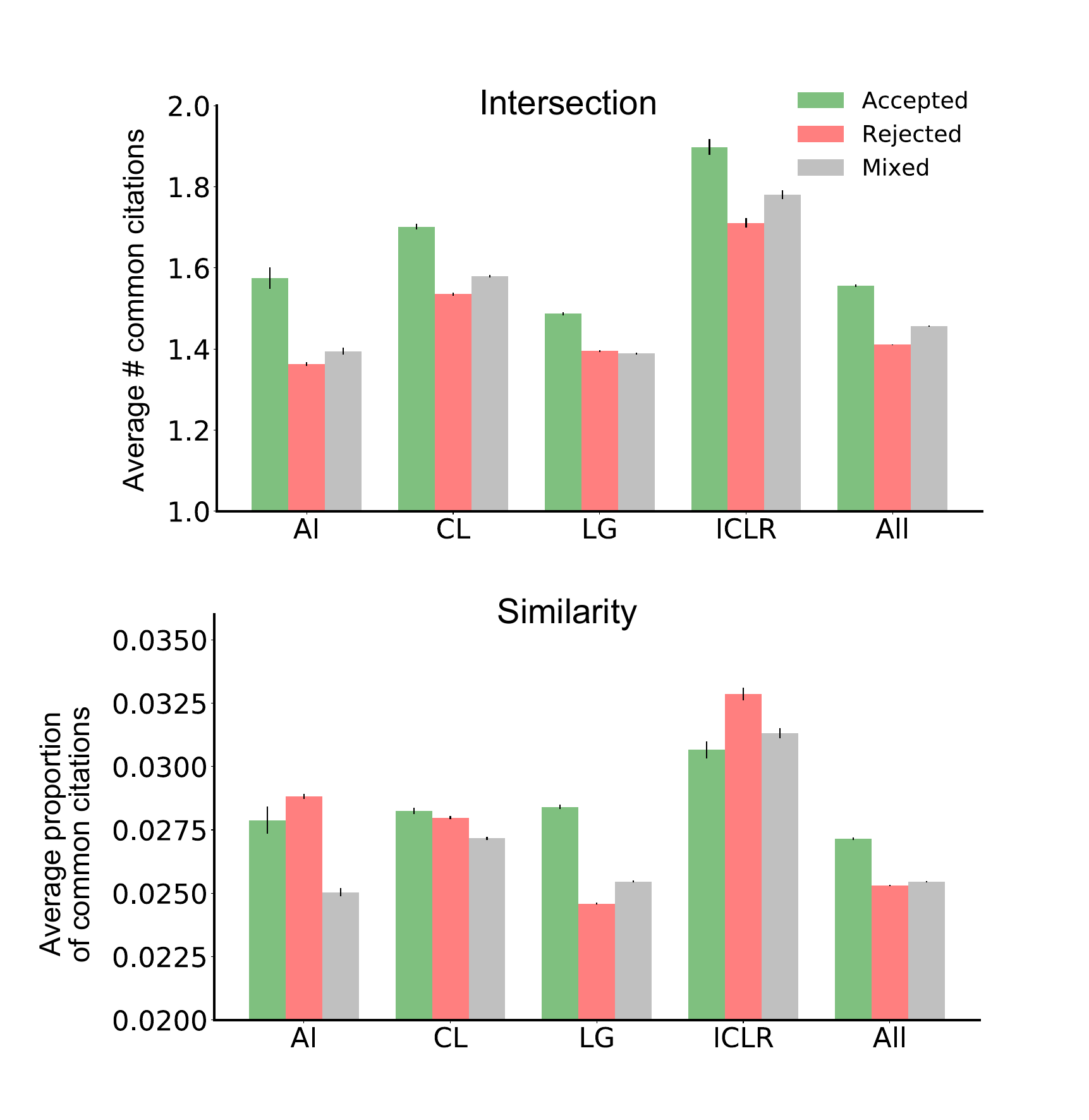}
\caption{\label{fig:PR_BC} Bibliographic coupling between accepted and rejected manuscripts.}
\end{figure}

We found that accepted manuscripts had more references in common (Fig. \ref{fig:PR_BC}) than the two other categories of manuscripts. The effect was slightly weaker for the Jaccard similarity (intersection over union of citations) and less consistent across platforms than the intersection. However, both metrics account for about 0.2\% of the variance (All platforms, Jaccard: 0.228\% and intersection: 0.21\%). 

\subsubsection{Semantic similarity and peer review outcome}
Having established that semantic similarity and bibliographic coupling capture different aspects of the relationship between documents, we also analysed the semantic similarity of the documents from the four platforms. Thus, for each platform we computed the td-idf distance between all pairs of document based on their word stem. 

\begin{figure}[H]
\centering
\includegraphics[width=0.65\textwidth]{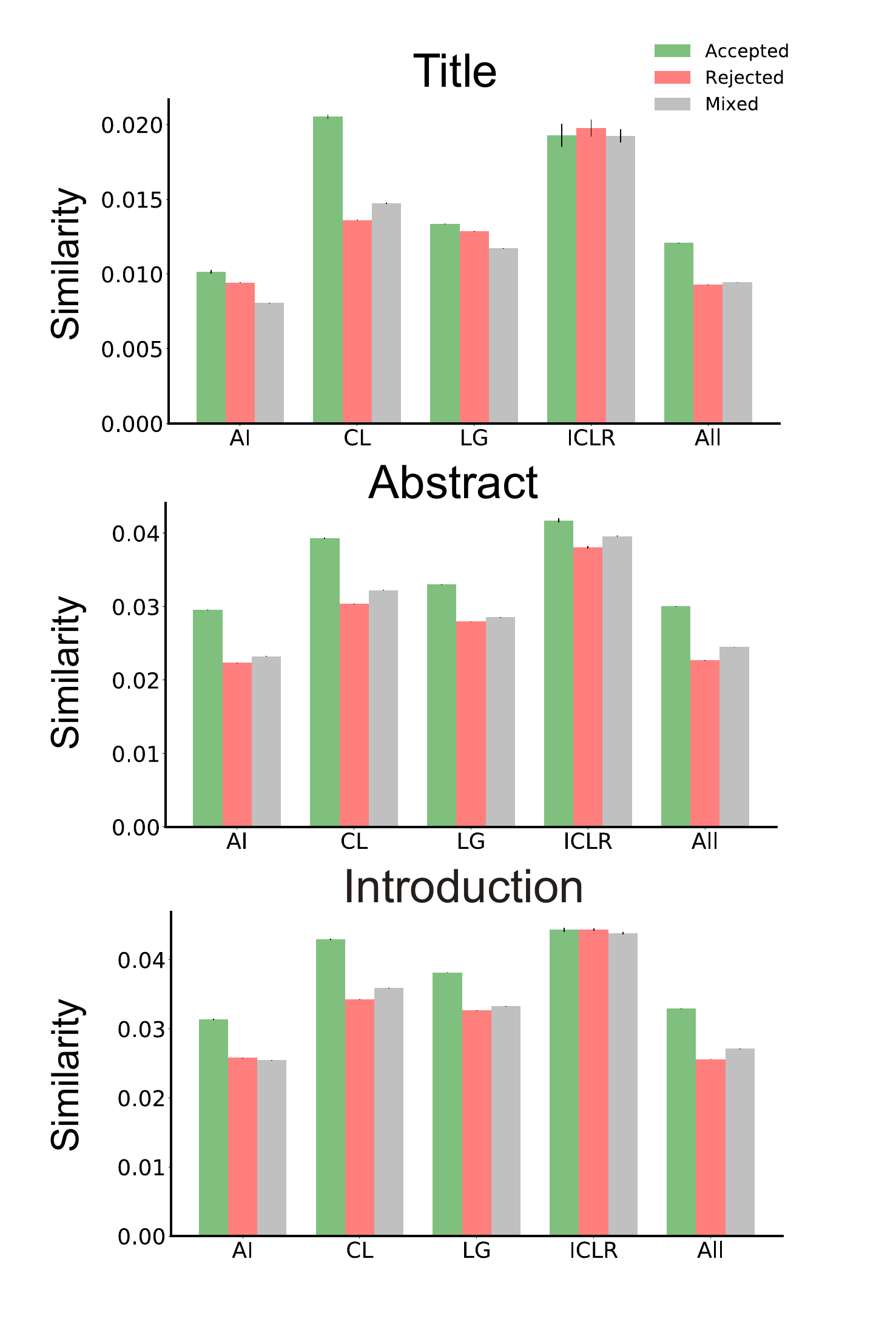}
\caption{\label{fig:PR_sim} Semantic similarity between accepted and rejected manuscripts.}
\end{figure}

Overall, we found that accepted manuscripts were more similar to each other than rejected manuscripts based on their abstracts and introduction (Fig.\ref{fig:PR_sim}). We found a stronger effect for the similarity between the introduction ($R^2 = 0.01$) than for the abstract ($R^2 = 0.006$) when all platforms were pooled together. In other words, accepted pairs of manuscripts were more similar to each other compared to the other two pair types.

This analysis of the semantic similarity of documents (for both citations and text) showed some high levels trends based on whether or not the manuscripts were accepted after peer review. We therefore next examined the text content of the manuscripts with a more detailed approach to gain more insights on the patterns uncovered by the analysis on bibliographic coupling and textual similarity.

\subsection{Words as a predictor of acceptance}
Finally, after having established high-level associations between the rejected and accepted manuscripts, we attempted to predict the peer review outcome with a logistic regression using a bag-of-words approach. Overall, the model was fairly successful at predicting the peer review outcome on a 10-fold cross-validated dataset (Tables \ref{title_stats},\ref{abstract_stats} \& \ref{introduction_stats}, random performance $\sim 0.5$ for all three metrics). The model was the most successful when the text of the introduction was used, followed by the text of the abstract and of the title. 

\begin{figure}[H]
\centering
\includegraphics[width=1\textwidth]{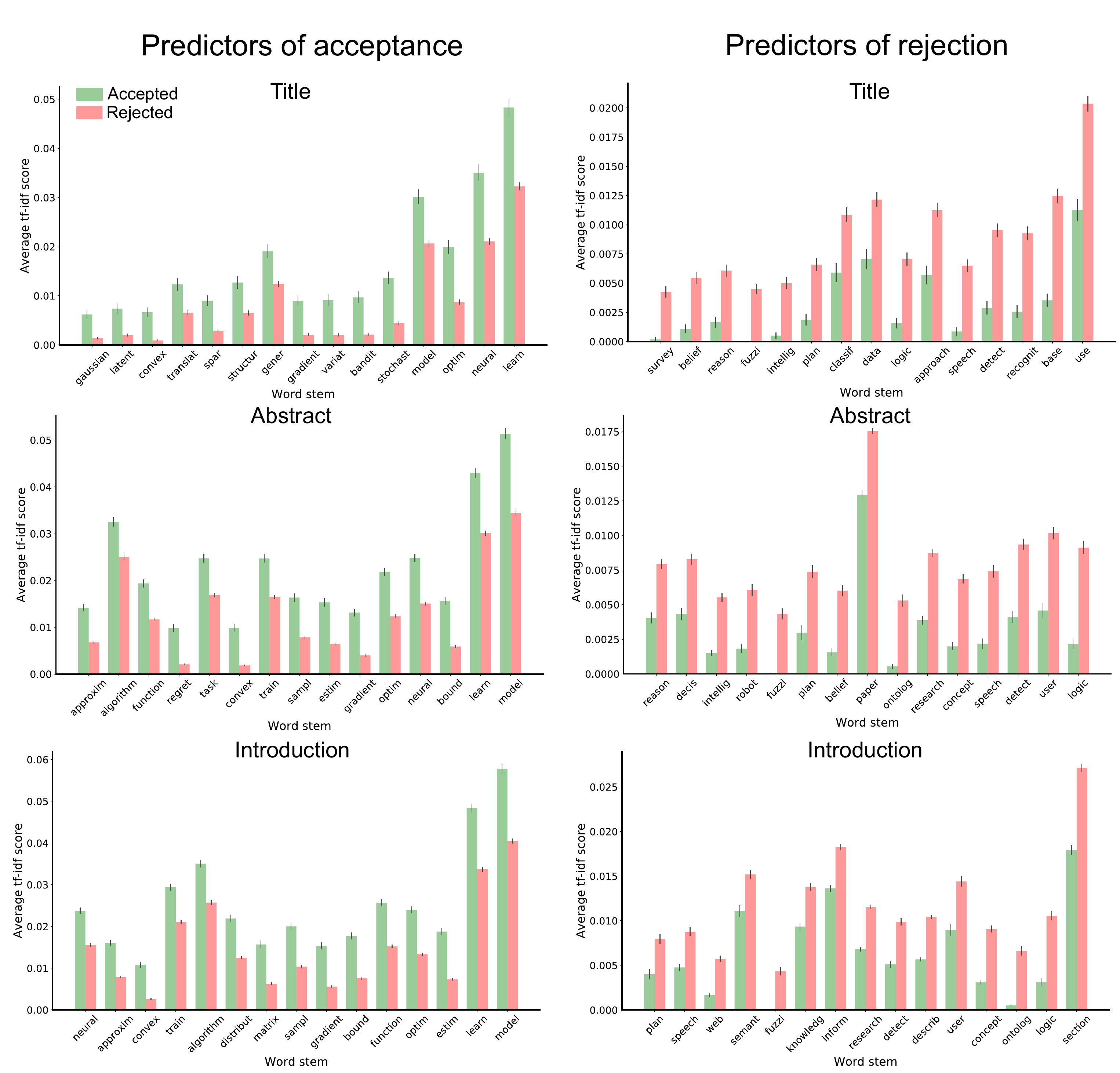}
\caption{\label{fig:Words_pred} Most important word stems for predicting peer review outcome.}
\end{figure}

After having established that we could predict to some extent the outcome of the peer review process with the word content of the manuscripts, we performed a more detailed analysis to try to get some insight about the key predictors of the outcome. We therefore computed the average tf-idf score of each stem for accepted and rejected manuscripts, and obtained measure of "importance" based on the difference between the two averages. This approach allowed us to identify the most important keywords predicting the acceptance of a manuscript (Fig. \ref{fig:Words_pred}).

Although some differences were noticeable across platforms regarding the predictors of acceptance (Tables \ref{title_words_acc}, \ref{abstract_words_acc} \& \ref{introduction_words_acc}) and rejection (Tables \ref{title_words_rej}, \ref{abstract_words_rej} \& \ref{introduction_words_rej}), some robust patterns emerged. Word stems related to sub fields of neural networks and machine learning (e.g., learn, neural, gradient, train) were increasing the odds of the manuscript to be accepted. However word stems related to the sub fields of logic, symbolic processing and knowledge representation (e.g, use, base, system, logic, fuzzi, knowledg, rule) were decreasing the odds that a manuscript would get accepted.

\begin{table*}
\centering
\begin{filecontents*}{data/title_stats.csv}
,Precision,Recall,F-score
AI,0.495 ± 0.048,0.501 ± 0.001,0.473 ± 0.002
CL,0.666 ± 0.029,0.538 ± 0.007,0.515 ± 0.011
LG,0.683 ± 0.01,0.629 ± 0.008,0.63 ± 0.009
ICLR,0.394 ± 0.045,0.486 ± 0.01,0.397 ± 0.026
All,0.694 ± 0.006,0.578 ± 0.003,0.58 ± 0.004
\end{filecontents*}
\csvautotabular{data/title_stats.csv}
\caption{Title based prediction performance (macro averaging).} \label{title_stats}
\begin{filecontents*}{data/abstract_stats.csv}
,Precision,Recall,F-score
AI,0.598 ± 0.077,0.506 ± 0.003,0.484 ± 0.007
CL,0.691 ± 0.025,0.553 ± 0.008,0.54 ± 0.013
LG,0.724 ± 0.006,0.674 ± 0.004,0.682 ± 0.005
ICLR,0.426 ± 0.057,0.494 ± 0.012,0.403 ± 0.02
All,0.734 ± 0.006,0.632 ± 0.005,0.65 ± 0.006
\end{filecontents*}
\csvautotabular{data/abstract_stats.csv}
\caption{Abstract based prediction performance (macro averaging).} \label{abstract_stats}
\begin{filecontents*}{data/introduction_stats.csv}
,Precision,Recall,F-score
AI,0.812 ± 0.035,0.523 ± 0.005,0.51 ± 0.01
CL,0.692 ± 0.015,0.57 ± 0.005,0.562 ± 0.007
LG,0.736 ± 0.009,0.712 ± 0.009,0.716 ± 0.009
ICLR,0.595 ± 0.064,0.527 ± 0.014,0.448 ± 0.026
All,0.744 ± 0.005,0.667 ± 0.005,0.684 ± 0.006
\end{filecontents*}
\csvautotabular{data/introduction_stats.csv} 
\caption{Introduction based prediction performance (macro averaging).} \label{introduction_stats}  
\end{table*}

\section{Discussion and Conclusion}
\subsection{Summary of results}
From a linguistic point of view, our results suggest that accepted manuscripts could be written in a more complex, less accessible English. Using two indices of readability, one of which is agnostic to the word content of the manuscript (FRE), we found that the accepted manuscripts obtained lower readability scores. Strikingly, we found the same effect for almost all our independent datasets. The same pattern was also observed for the title, the abstract and the introduction. Using a different type of readability indicator - the proportion of general, specific scientific and AI jargon words - we found that manuscripts that contained a greater proportion of jargon words were more likely to be accepted. This finding may partly explain that the readability of manuscripts has steadily declined during the last century \cite{plaven-sigray_readability_2017}. In other words, it is possible that part of this decline could be attributed to a selection process taking place during peer review. 

When considering the word content of the accepted manuscripts, we found that they had words that were acquired at a later age, that were more abstract and that were less common than the words from the rejected manuscripts. Additionally, these manuscripts were shorter and had increased lexical diversity. Once again, the effect size were small given the highly multivariate determination of the peer review outcome. The effects were replicated across multiples independent datasets from different fields in AI, which strengthens the conclusions of our analysis. 

From a content point of view, we compared manuscripts based on their referencing patterns and word content. We compared the coupling both based on the raw number of common references (intersection) and the fraction of overlap between the manuscripts' references (Jaccard similarity). We found that accepted pairs had a larger intersection than other pairs, and found a similar, but less reliable effect for the similarity. We used a tf-idf vectorial representation of the text from all manuscripts in the database, compared all possible pairs of manuscripts, and we found that pairs of accepted manuscript had considerably more overlap between their word content. This high level analysis of the manuscript's content revealed that some topics might be associated with different odds of acceptance. We performed a correlation between the bibliographic coupling and the semantic similarity to get an idea of the how independent was the information provided by these two semantic indicators. As reported previously \cite{sainte-marie_you_2018}, we found a weak to moderate correlation between the two, which suggest that they provide distinct sources of information about topic similarity in accepted manuscripts. 

Finally, we built a logistic regression to predict the peer review outcome, which revealed that using the title, abstract or introduction words lead to robust predictions. Our results are compatible with the presence of content bias, where trending topics in AI such as machine learning and neural networks were linked with greater acceptance rate, whereas words related to symbolic processing and knowledge-based reasoning lead to lower acceptance rates. 

\subsection{Implications}

Taken together, our analysis of the linguistic aspects of manuscripts are coherent with linguistic biases during peer review. It has been reported that writers using English as their main language (L1) use words that are more abstract and less frequent than writers with English as their second language (L2) \cite{crossley_computational_2009}. Additionally, this effect is exacerbated by the L2 proficiency (where larger differences are observed for beginners than advanced L2 speakers) \cite{crossley_predicting_2011}. The complexity of L2 writing was also shown to correlate with proficiency \cite{kim_predicting_2014,lahuerta_martinez_analysis_2018,radhiah_longitudinal_2018}. Our results are therefore compatible with the hypothesis that L2 writers are less likely to get their manuscript accepted at peer review.

Our results are also compatible with a content bias where manuscripts on the topics of machine learning techniques and neural networks have greater odds to be accepted at peer review. Leading figures of the AI community have raised their voice against the overwhelming dominance of neural networks and deep learning in the domain of AI \cite{marcus_deep_2018,jordan_artificial_2018,knight_one_2018}. Recent successes of deep learning and neural networks might explain their dominance in the field, but a bias against other techniques might impede developments similar to the ones that lead to the breakthroughs underlying the deep learning revolution \cite{krizhevsky_imagenet_2012}. Following this idea, several researchers have indicated that symbolic processing could hold the answer to shortcomings of deep learning \cite{geffner_model-free_2018,marcus_deep_2018,garnelo_reconciling_2019}. 

Other than biases during peer review, our results have implications for the quality of scientific communications. A recent report on reproducibility in machine learning found a positive correlation between the readability of a manuscript and successful replication of the claims that it makes \cite{raff_step_2019}. Selecting for less readable manuscripts during peer review may therefore increase the proportion of non reproducible research findings.

\subsection{Limitations}
Although the main objective of our analysis was to investigate the presence of content or linguistic biases in peer review, all our of analysis are correlational, and there are possible confounds that could explain our results. For instance, while our findings that some linguistic aspects of the manuscripts - the readability and psycholinguistic attributes - were correlated with the peer review outcome, we cannot infer that those variables directly lowered the odds of acceptance.

Similarly, we cannot infer that there is a bias against manuscripts on the topic of machine learning techniques and neural networks. For instance, reviewers favouring high benchmark performance might accept more manuscripts using the state of the art techniques. In the scenario, reviewers would not reject a manuscript using a non-mainstream technique because of bias against it, but simply because they value some aspects where it under performs. 

Another limitation to our findings is the methodology of the peer read dataset \cite{kang_dataset_2018}. For most manuscripts included in the dataset, their status is inferred and the true outcome of the peer review process is unknown. Although \cite{kang_dataset_2018} validated their method on a subset of their data, the accuracy is not perfect. However, we believe that the large size of the dataset is enough to counteract this source of noise. Only the minority of manuscripts included in their dataset had a true peer review outcome provided by the publishing venue. This highlight the need for publishers and conferences to open their peer review process in order to further advance our understanding of the strengths and limitations of the peer review process.

In sum, our results are compatible with the presence of a linguistic and a content bias in the peer review process of major conferences in AI. Although we were able to replicate our results across different dataset, similar studies have to be conducted both in the field of AI and in other disciplines to validate the conclusions of our study. 

\section*{Acknowlegements}{This research was funded by the Canada Research Chair program and the Social Sciences and Humanities Research Council of Canada.}

\bibliographystyle{ieeetr}
\bibliography{PeerRead}


\afterpage{%
    \clearpage
    \begin{table*}
	    
        \centering 
        \resizebox{\textwidth}{!}{
		\begin{filecontents*}{data/Read_US.csv}
Section,Variable,p,Acc-Rej,Mean rejected,S.d. rejected,N rejected,Mean accepted,S.d. accepted,N accepted,Cohen's d
Title,FRE,2.9E-02,-3.56,-7.89,40.84,1714,-11.45,43.08,1113,0.085
Abstract,FRE,1.7E-07,-2.35,18.47,12.69,1882,16.12,11.94,1239,0.190
Introduction,FRE,3.0E-01,-0.38,27.14,10.04,1545,26.75,9.22,1215,0.040
Title,NDC,9.8E-09,0.53,15.19,2.38,1714,15.72,2.38,1113,0.222
Abstract,NDC,5.7E-06,0.15,12.50,0.90,1882,12.65,0.87,1239,0.165
Introduction,NDC,5.7E-11,0.18,12.12,0.77,1545,12.30,0.69,1215,0.249
Title,General,1.2E-02,0.02,0.38,0.22,1715,0.41,0.23,1113,0.099
Abstract,General,5.6E-06,0.01,0.30,0.07,1882,0.31,0.06,1239,0.163
Introduction,General,1.7E-01,0.00,0.26,0.06,1545,0.26,0.05,1215,0.052
Title,Specific,1.2E-02,0.02,0.41,0.23,1715,0.43,0.24,1113,0.098
Abstract,Specific,1.1E-05,0.01,0.32,0.07,1882,0.33,0.07,1239,0.158
Introduction,Specific,1.1E-04,0.01,0.28,0.06,1545,0.29,0.05,1215,0.147
Title,AI,2.9E-02,0.01,0.09,0.13,1715,0.10,0.14,1113,0.085
Abstract,AI,1.1E-05,0.01,0.05,0.03,1882,0.05,0.03,1239,0.159
Introduction,AI,1.0E-06,0.00,0.04,0.02,1545,0.04,0.02,1215,0.186
\end{filecontents*}
\csvautotabular{data/Read_US.csv}
		}
		\caption{Summary of statistics for readability and jargon usage of US based accepted and rejected manuscripts.} 
		\label{Read_US}
    \end{table*}
    
    \begin{table*}
	    
        \centering 
        \resizebox{\textwidth}{!}{
		\begin{filecontents*}{data/Read_non_US.csv}
Section,Variable,p,Acc-Rej,Mean rejected,S.d. rejected,N rejected,Mean accepted,S.d. accepted,N accepted,Cohen's d
Title,FRE,6.3E-06,-5.11,-4.52,38.52,6257,-9.62,41.16,1635,0.131
Abstract,FRE,2.4E-10,-2.10,20.16,13.59,7100,18.06,12.25,1817,0.157
Introduction,FRE,1.7E-01,0.36,28.02,10.52,5514,28.38,9.29,1769,0.036
Title,NDC,4.6E-20,0.61,15.01,2.37,6257,15.61,2.36,1635,0.256
Abstract,NDC,2.0E-05,0.11,12.44,1.09,7100,12.54,0.91,1817,0.101
Introduction,NDC,2.0E-16,0.18,12.05,1.03,5514,12.23,0.71,1769,0.187
Title,General,8.3E-04,0.02,0.38,0.22,6262,0.40,0.24,1635,0.098
Abstract,General,4.1E-26,0.02,0.29,0.07,7104,0.31,0.07,1817,0.262
Introduction,General,8.0E-10,0.01,0.25,0.06,5514,0.26,0.05,1769,0.150
Title,Specific,8.1E-05,0.03,0.41,0.23,6262,0.44,0.25,1635,0.116
Abstract,Specific,1.4E-26,0.02,0.31,0.08,7104,0.33,0.07,1817,0.263
Introduction,Specific,1.3E-35,0.02,0.27,0.06,5514,0.29,0.05,1769,0.310
Title,AI,1.5E-03,0.01,0.09,0.13,6262,0.10,0.15,1635,0.095
Abstract,AI,8.0E-37,0.01,0.04,0.03,7104,0.05,0.03,1817,0.330
Introduction,AI,3.0E-54,0.01,0.03,0.02,5514,0.04,0.02,1769,0.400
\end{filecontents*}
\csvautotabular{data/Read_non_US.csv}
		}
		\caption{Summary of statistics for readability and jargon usage of non-US based accepted and rejected manuscripts.} 
		\label{Read_non_US}
    \end{table*}    
}

\afterpage{
\begin{table*}
\centering 
\resizebox{\textwidth}{!}{
\begin{filecontents*}{data/Linguistic.csv}
Section,Variable,p,Acc-Rej,Mean rejected,S.d. rejected,N rejected,Mean accepted,S.d. accepted,N accepted,Cohen's d
Title,Frequency tokens,8.12E-48,-0.21,3.12,0.60,7959,2.91,0.66,2742,0.34
Abstract,Frequency tokens,1.37E-27,-0.04,3.87,0.21,8983,3.83,0.19,3056,0.22
Introduction,Frequency tokens,7.01E-27,-0.04,3.95,0.18,7059,3.91,0.14,2984,0.21
Title,AOA tokens,4.64E-20,0.36,7.85,1.57,7920,8.21,1.80,2710,0.22
Abstract,AOA tokens,3.79E-20,0.08,6.50,0.44,8983,6.58,0.39,3056,0.18
Introduction,AOA tokens,5.19E-19,0.06,6.32,0.35,7059,6.38,0.27,2984,0.18
Title,Concreteness tokens,5.91E-02,-0.02,2.54,0.38,7953,2.53,0.36,2736,0.04
Abstract,Concreteness tokens,4.13E-01,0.00,2.36,0.12,8983,2.35,0.10,3056,0.02
Introduction,Concreteness tokens,1.04E-06,-0.01,2.33,0.09,7059,2.32,0.07,2984,0.10
Title,Frequency types,2.67E-44,-0.20,3.10,0.59,7959,2.90,0.66,2742,0.33
Abstract,Frequency types,5.08E-11,-0.03,3.40,0.21,8983,3.38,0.18,3056,0.13
Introduction,Frequency types,6.30E-18,-0.03,3.18,0.19,7059,3.15,0.14,2984,0.17
Title,AOA types,1.45E-18,0.34,7.87,1.56,7920,8.22,1.80,2710,0.21
Abstract,AOA types,2.92E-12,0.06,7.16,0.46,8983,7.22,0.42,3056,0.14
Introduction,AOA types,1.40E-17,0.06,7.39,0.38,7059,7.45,0.31,2984,0.17
Title,Concreteness types,1.68E-02,-0.02,2.55,0.37,7953,2.53,0.35,2736,0.05
Abstract,Concreteness types,2.66E-15,-0.02,2.50,0.12,8983,2.48,0.10,3056,0.15
Introduction,Concreteness types,1.47E-20,-0.02,2.55,0.09,7059,2.53,0.07,2984,0.18
Title,\# tokens,2.28E-60,-1.27,9.78,5.03,7977,8.51,2.73,2748,0.28
Abstract,\# tokens,3.93E-03,-10.82,169.34,202.13,8986,158.52,170.54,3056,0.06
Introduction,\# tokens,5.39E-08,-98.57,851.91,1351.31,7059,753.34,455.07,2984,0.09
Title,\# types,5.39E-67,-1.15,9.60,3.80,7977,8.45,2.63,2748,0.33
Abstract,\# types,3.93E-03,-3.52,103.61,61.04,8986,100.09,57.27,3056,0.06
Introduction,\# types,3.85E-03,-10.23,349.60,218.87,7059,339.37,130.78,2984,0.05
Title,TTR,1.34E-23,0.01,0.99,0.04,7977,1.00,0.02,2748,0.18
Abstract,TTR,2.76E-31,0.02,0.66,0.08,8986,0.68,0.08,3056,0.24
Introduction,TTR,9.02E-04,0.01,0.48,0.09,7059,0.49,0.08,2984,0.07
\end{filecontents*}
\csvautotabular{data/Linguistic.csv}}
\caption{Summary of statistics for lexical features for accepted and rejected manuscripts.} 
\label{Linguistic}
\end{table*}}

\afterpage{%
    \clearpage
    \thispagestyle{empty}
    \begin{landscape}
    \begin{table*}
        \centering 
        \resizebox{\columnwidth}{170pt}{
		\begin{filecontents*}{data/title_words_acc.csv}
,AI Stem,AI Score accepted,AI Score rejected,CL Stem,CL Score accepted,CL Score rejected,LG Stem,LG Score accepted,LG Score rejected,ICLR Stem,ICLR Score accepted,ICLR Score rejected,All Stem,All Score accepted,All Score rejected
0,model,0.0345 ± 0.0043,0.0201 ± 0.0012,learn,0.0538 ± 0.0039,0.0452 ± 0.002,model,0.0252 ± 0.0018,0.019 ± 0.0012,featur,0.0231 ± 0.008,0.0017 ± 0.0017,learn,0.0483 ± 0.0017,0.0323 ± 0.0008
1,deep,0.0322 ± 0.0047,0.0215 ± 0.0013,bandit,0.0131 ± 0.003,0.0058 ± 0.0012,cluster,0.0124 ± 0.0017,0.0085 ± 0.001,structur,0.0234 ± 0.0086,0.0023 ± 0.0023,neural,0.035 ± 0.0017,0.0211 ± 0.0007
2,recognit,0.0136 ± 0.0035,0.0069 ± 0.0008,deep,0.0205 ± 0.0031,0.0141 ± 0.0015,graph,0.009 ± 0.0014,0.0057 ± 0.0008,latent,0.0198 ± 0.0088,0.0 ± 0.0,optim,0.0199 ± 0.0015,0.0087 ± 0.0005
3,visual,0.0097 ± 0.0028,0.0035 ± 0.0006,algorithm,0.0212 ± 0.0031,0.0161 ± 0.0016,languag,0.0077 ± 0.0012,0.0045 ± 0.0007,adapt,0.0205 ± 0.0082,0.0022 ± 0.0022,model,0.0302 ± 0.0015,0.0207 ± 0.0007
4,machin,0.0168 ± 0.0034,0.011 ± 0.001,linear,0.0127 ± 0.0029,0.0077 ± 0.0013,featur,0.0136 ± 0.0016,0.0105 ± 0.0011,stochast,0.0203 ± 0.0075,0.0042 ± 0.003,stochast,0.0136 ± 0.0013,0.0044 ± 0.0004
5,data,0.0175 ± 0.0034,0.0127 ± 0.001,fast,0.0101 ± 0.0025,0.0055 ± 0.001,gradient,0.0086 ± 0.0014,0.0057 ± 0.0008,learn,0.0695 ± 0.0094,0.0555 ± 0.0079,bandit,0.0097 ± 0.0012,0.0021 ± 0.0003
6,oneshot,0.0051 ± 0.003,0.0004 ± 0.0003,space,0.0064 ± 0.0019,0.0019 ± 0.0006,bayesian,0.0104 ± 0.0014,0.0075 ± 0.001,machin,0.024 ± 0.0086,0.0102 ± 0.0052,variat,0.0091 ± 0.0012,0.0021 ± 0.0003
7,measur,0.0079 ± 0.0026,0.0035 ± 0.0007,object,0.0069 ± 0.0021,0.0025 ± 0.0007,onlin,0.0138 ± 0.0017,0.011 ± 0.0011,svm,0.0136 ± 0.0078,0.0 ± 0.0,gradient,0.009 ± 0.0011,0.0021 ± 0.0003
8,align,0.0048 ± 0.0024,0.0006 ± 0.0003,onlin,0.0158 ± 0.0029,0.0115 ± 0.0015,function,0.0078 ± 0.0013,0.0052 ± 0.0008,parallel,0.0126 ± 0.0073,0.0 ± 0.0,gener,0.0191 ± 0.0014,0.0124 ± 0.0006
9,sequenti,0.0068 ± 0.0031,0.0026 ± 0.0006,entropi,0.0042 ± 0.0019,0.0004 ± 0.0003,vector,0.005 ± 0.001,0.0025 ± 0.0006,graph,0.0223 ± 0.0092,0.0098 ± 0.0049,structur,0.0127 ± 0.0013,0.0065 ± 0.0005
10,neural,0.0266 ± 0.0041,0.0225 ± 0.0013,pca,0.0057 ± 0.0023,0.0019 ± 0.0007,optim,0.0184 ± 0.0018,0.016 ± 0.0012,exponenti,0.0124 ± 0.0072,0.0 ± 0.0,spar,0.009 ± 0.0011,0.0029 ± 0.0004
11,entiti,0.0053 ± 0.0021,0.0011 ± 0.0004,regret,0.0061 ± 0.0022,0.0024 ± 0.0008,exploit,0.0032 ± 0.0008,0.0008 ± 0.0003,convolut,0.0209 ± 0.0084,0.0093 ± 0.0046,translat,0.0123 ± 0.0013,0.0066 ± 0.0005
12,relat,0.0084 ± 0.0028,0.0045 ± 0.0007,comparison,0.0061 ± 0.002,0.0024 ± 0.0007,kmean,0.0041 ± 0.0011,0.0018 ± 0.0005,extrem,0.0112 ± 0.0064,0.0 ± 0.0,convex,0.0066 ± 0.001,0.0009 ± 0.0002
13,disambigu,0.0046 ± 0.0023,0.0006 ± 0.0003,crowd,0.004 ± 0.002,0.0003 ± 0.0003,memori,0.0052 ± 0.0011,0.0029 ± 0.0006,duel,0.0104 ± 0.0077,0.0 ± 0.0,latent,0.0074 ± 0.001,0.002 ± 0.0003
14,latent,0.0084 ± 0.0028,0.0044 ± 0.0007,reinforc,0.0105 ± 0.0025,0.0068 ± 0.0012,depend,0.0046 ± 0.001,0.0023 ± 0.0006,semisupervi,0.013 ± 0.0075,0.0027 ± 0.0027,gaussian,0.0062 ± 0.0009,0.0014 ± 0.0003
15,featur,0.0158 ± 0.0036,0.0119 ± 0.0011,constraint,0.0065 ± 0.0022,0.003 ± 0.0008,spar,0.0089 ± 0.0014,0.0067 ± 0.0009,ensembl,0.0102 ± 0.0074,0.0 ± 0.0,par,0.0069 ± 0.001,0.0023 ± 0.0003
16,squar,0.0048 ± 0.0024,0.001 ± 0.0004,rule,0.0054 ± 0.0018,0.0019 ± 0.0006,regret,0.0043 ± 0.001,0.0023 ± 0.0006,denoi,0.0102 ± 0.0072,0.0 ± 0.0,infer,0.0108 ± 0.0011,0.0065 ± 0.0005
17,gener,0.0158 ± 0.0036,0.012 ± 0.001,nonconvex,0.0048 ± 0.002,0.0013 ± 0.0006,distribut,0.0105 ± 0.0015,0.0086 ± 0.001,boltzmann,0.0101 ± 0.0071,0.0 ± 0.0,matrix,0.0061 ± 0.0009,0.002 ± 0.0003
18,discret,0.006 ± 0.0027,0.0023 ± 0.0005,pairwi,0.004 ± 0.0018,0.0006 ± 0.0004,selfpac,0.0018 ± 0.0008,0.0 ± 0.0,advanc,0.0097 ± 0.007,0.0 ± 0.0,regret,0.0048 ± 0.0009,0.0007 ± 0.0002
19,speech,0.0073 ± 0.0026,0.0036 ± 0.0007,music,0.0049 ± 0.0021,0.0016 ± 0.0006,submodular,0.0038 ± 0.0011,0.002 ± 0.0006,bootstrap,0.0097 ± 0.0068,0.0 ± 0.0,recurr,0.0112 ± 0.0013,0.0071 ± 0.0005
20,sensor,0.0047 ± 0.0021,0.0011 ± 0.0004,featur,0.0163 ± 0.0029,0.013 ± 0.0015,parallel,0.0042 ± 0.001,0.0024 ± 0.0006,traffic,0.0095 ± 0.0068,0.0 ± 0.0,reinforc,0.0109 ± 0.0012,0.0069 ± 0.0005
21,supervi,0.0076 ± 0.0026,0.004 ± 0.0007,predict,0.0136 ± 0.0026,0.0104 ± 0.0013,chine,0.0018 ± 0.0007,1e-04 ± 1e-04,smooth,0.0093 ± 0.0066,0.0 ± 0.0,fast,0.0069 ± 0.0009,0.0031 ± 0.0004
22,largesc,0.0074 ± 0.0026,0.0038 ± 0.0007,convex,0.0093 ± 0.0027,0.0061 ± 0.0012,exponenti,0.0028 ± 0.0009,0.0012 ± 0.0004,pca,0.0091 ± 0.0065,0.0 ± 0.0,sampl,0.007 ± 0.001,0.0032 ± 0.0004
23,pursuit,0.0039 ± 0.0022,0.0004 ± 0.0002,face,0.0034 ± 0.0016,0.0003 ± 0.0003,hierarch,0.0059 ± 0.0012,0.0043 ± 0.0007,distribut,0.013 ± 0.0065,0.0039 ± 0.0028,network,0.0307 ± 0.0016,0.0269 ± 0.0008
24,addit,0.0045 ± 0.0026,0.001 ± 0.0004,squar,0.0036 ± 0.0018,0.0005 ± 0.0003,support,0.0041 ± 0.001,0.0025 ± 0.0006,errorcorrect,0.0088 ± 0.0088,0.0 ± 0.0,linear,0.0073 ± 0.001,0.0036 ± 0.0004
25,multimod,0.0054 ± 0.0024,0.0019 ± 0.0005,markov,0.0058 ± 0.002,0.0027 ± 0.0008,review,0.0029 ± 0.0008,0.0013 ± 0.0004,transform,0.0126 ± 0.0072,0.0038 ± 0.0038,bound,0.0069 ± 0.0009,0.0032 ± 0.0004
26,hierarch,0.0082 ± 0.0029,0.0047 ± 0.0008,multiarm,0.0037 ± 0.0019,0.0006 ± 0.0004,descent,0.005 ± 0.0011,0.0035 ± 0.0007,fast,0.0173 ± 0.0076,0.0088 ± 0.0046,minim,0.0054 ± 0.0008,0.0017 ± 0.0003
27,miss,0.0036 ± 0.0021,0.0002 ± 0.0002,limit,0.0049 ± 0.002,0.0018 ± 0.0007,smooth,0.0029 ± 0.0009,0.0014 ± 0.0005,tempor,0.0084 ± 0.006,0.0 ± 0.0,kernel,0.0074 ± 0.0011,0.0037 ± 0.0004
28,backpropag,0.0042 ± 0.0024,0.0008 ± 0.0003,use,0.0187 ± 0.0026,0.0156 ± 0.0013,theoret,0.0025 ± 0.0008,0.0011 ± 0.0004,decod,0.0084 ± 0.006,0.0 ± 0.0,sequenc,0.0073 ± 0.0011,0.0037 ± 0.0004
29,biolog,0.0036 ± 0.0022,0.0003 ± 0.0002,lifelong,0.0034 ± 0.002,0.0003 ± 0.0003,reinforc,0.0093 ± 0.0015,0.0079 ± 0.0009,approach,0.009 ± 0.0053,0.0006 ± 0.0006,emb,0.0106 ± 0.0011,0.007 ± 0.0005
30,annot,0.0041 ± 0.0021,0.0008 ± 0.0003,compon,0.0042 ± 0.0019,0.0012 ± 0.0005,sparsiti,0.0023 ± 0.0008,0.0008 ± 0.0004,consist,0.0083 ± 0.0058,0.0 ± 0.0,nonparametr,0.0044 ± 0.0009,0.0008 ± 0.0002
31,topic,0.0056 ± 0.0023,0.0023 ± 0.0005,track,0.0033 ± 0.0015,0.0003 ± 0.0002,percept,0.0018 ± 0.0007,0.0004 ± 0.0002,smile,0.0082 ± 0.0082,0.0 ± 0.0,attent,0.0068 ± 0.001,0.0033 ± 0.0004
32,predict,0.0151 ± 0.0036,0.0118 ± 0.001,propag,0.0047 ± 0.002,0.0017 ± 0.0007,provabl,0.002 ± 0.0008,0.0006 ± 0.0003,cluster,0.0127 ± 0.0063,0.0047 ± 0.0033,word,0.0106 ± 0.0011,0.0071 ± 0.0005
33,mixabl,0.0032 ± 0.0023,0.0 ± 0.0,hyperparamet,0.0034 ± 0.002,0.0005 ± 0.0004,transfer,0.0043 ± 0.0011,0.0029 ± 0.0006,input,0.008 ± 0.0057,0.0 ± 0.0,estim,0.0076 ± 0.001,0.0041 ± 0.0004
34,hypergraph,0.0033 ± 0.0024,0.0002 ± 0.0002,effici,0.0111 ± 0.0025,0.0082 ± 0.0013,thompson,0.0019 ± 0.0007,0.0005 ± 0.0003,interfac,0.0079 ± 0.0056,0.0 ± 0.0,predict,0.0121 ± 0.0012,0.0086 ± 0.0005
35,variabl,0.0053 ± 0.0024,0.0021 ± 0.0005,generali,0.0031 ± 0.0015,0.0002 ± 0.0002,speaker,0.0023 ± 0.0008,0.001 ± 0.0004,onlin,0.0248 ± 0.0086,0.0171 ± 0.006,converg,0.0047 ± 0.0008,0.0012 ± 0.0003
36,depend,0.0063 ± 0.0024,0.0031 ± 0.0006,model,0.0241 ± 0.003,0.0213 ± 0.0016,prior,0.0021 ± 0.0007,0.0008 ± 0.0003,exampl,0.0076 ± 0.0054,0.0 ± 0.0,random,0.0065 ± 0.0009,0.0031 ± 0.0004
37,librari,0.0038 ± 0.0023,0.0008 ± 0.0003,exploit,0.0043 ± 0.0018,0.0015 ± 0.0006,rel,0.002 ± 0.0008,0.0006 ± 0.0003,origami,0.0076 ± 0.0076,0.0 ± 0.0,spectral,0.0043 ± 0.0008,0.0009 ± 0.0002
38,perspect,0.0047 ± 0.0024,0.0016 ± 0.0005,valu,0.0035 ± 0.0016,0.0007 ± 0.0004,point,0.0031 ± 0.0009,0.0018 ± 0.0005,simplex,0.0075 ± 0.0075,0.0 ± 0.0,onlin,0.0102 ± 0.0011,0.0069 ± 0.0005
39,tree,0.0076 ± 0.0029,0.0046 ± 0.0007,statist,0.0062 ± 0.0021,0.0035 ± 0.0008,converg,0.0045 ± 0.001,0.0032 ± 0.0007,paint,0.0074 ± 0.0052,0.0 ± 0.0,nonconvex,0.0036 ± 0.0008,0.0003 ± 1e-04
40,multitask,0.0063 ± 0.0026,0.0032 ± 0.0007,sampl,0.007 ± 0.0022,0.0043 ± 0.001,larg,0.0041 ± 0.0009,0.0028 ± 0.0006,layerspecif,0.0072 ± 0.0072,0.0 ± 0.0,activ,0.0069 ± 0.001,0.0036 ± 0.0004
41,person,0.0044 ± 0.0023,0.0014 ± 0.0004,market,0.0029 ± 0.0015,0.0002 ± 0.0002,effici,0.0099 ± 0.0014,0.0086 ± 0.001,path,0.0072 ± 0.0072,0.0 ± 0.0,descent,0.0045 ± 0.0008,0.0013 ± 0.0002
42,classif,0.0153 ± 0.0034,0.0123 ± 0.0011,better,0.0033 ± 0.0017,0.0006 ± 0.0003,twitter,0.0016 ± 0.0006,0.0003 ± 0.0002,dictionari,0.0098 ± 0.0069,0.0027 ± 0.0027,hierarch,0.0065 ± 0.001,0.0034 ± 0.0004
43,consist,0.0043 ± 0.002,0.0013 ± 0.0004,activ,0.0062 ± 0.0021,0.0036 ± 0.0009,forest,0.0026 ± 0.0009,0.0013 ± 0.0005,graphlet,0.007 ± 0.007,0.0 ± 0.0,effici,0.0087 ± 0.001,0.0057 ± 0.0005
44,loop,0.0031 ± 0.0022,0.0002 ± 0.0002,mont,0.0032 ± 0.0017,0.0007 ± 0.0004,translat,0.0036 ± 0.0009,0.0023 ± 0.0006,consensusba,0.0069 ± 0.0069,0.0 ± 0.0,regular,0.0058 ± 0.0009,0.0028 ± 0.0004
45,nonparametr,0.0058 ± 0.0029,0.0029 ± 0.0007,carlo,0.0032 ± 0.0017,0.0007 ± 0.0004,queri,0.0023 ± 0.0008,0.001 ± 0.0004,train,0.0116 ± 0.0067,0.0047 ± 0.0034,submodular,0.0037 ± 0.0008,0.0009 ± 0.0002
46,varianc,0.0037 ± 0.0021,0.0008 ± 0.0003,coher,0.0029 ± 0.0017,0.0004 ± 0.0003,nearoptim,0.0019 ± 0.0008,0.0007 ± 0.0003,pragmat,0.0068 ± 0.0068,0.0 ± 0.0,depend,0.0051 ± 0.0008,0.0022 ± 0.0003
47,coevolut,0.0029 ± 0.0021,0.0 ± 0.0,contextu,0.004 ± 0.0018,0.0015 ± 0.0006,infer,0.0073 ± 0.0011,0.006 ± 0.0008,dropout,0.0068 ± 0.0068,0.0 ± 0.0,multitask,0.0047 ± 0.0009,0.0019 ± 0.0003
48,email,0.0028 ± 0.0016,0.0 ± 0.0,detect,0.0094 ± 0.0022,0.0069 ± 0.0011,semisupervi,0.0055 ± 0.0012,0.0042 ± 0.0008,nonadapt,0.0067 ± 0.0067,0.0 ± 0.0,match,0.0048 ± 0.0009,0.0019 ± 0.0003
49,genet,0.0037 ± 0.0022,0.001 ± 0.0004,estim,0.01 ± 0.0024,0.0076 ± 0.0012,detect,0.0086 ± 0.0012,0.0074 ± 0.0009,hidden,0.009 ± 0.0064,0.0025 ± 0.0025,regress,0.005 ± 0.0008,0.0022 ± 0.0003
\end{filecontents*}
\csvautotabular{data/title_words_acc.csv}
		}
		\caption{Most important predictors for acceptance in decreasing order of importance: title} \label{title_words_acc}
    \end{table*}
    \end{landscape}
    \clearpage
}

\afterpage{%
    \clearpage
    \thispagestyle{empty}
    \begin{landscape}
    \begin{table*}
        \centering 
        \resizebox{\columnwidth}{170pt}{
		\begin{filecontents*}{data/abstract_words_acc.csv}
,AI Stem,AI Score accepted,AI Score rejected,CL Stem,CL Score accepted,CL Score rejected,LG Stem,LG Score accepted,LG Score rejected,ICLR Stem,ICLR Score accepted,ICLR Score rejected,All Stem,All Score accepted,All Score rejected
0,measur,0.016 ± 0.0028,0.008 ± 0.0005,matrix,0.0167 ± 0.0027,0.0091 ± 0.001,word,0.0122 ± 0.0012,0.0078 ± 0.0007,algorithm,0.0446 ± 0.0049,0.024 ± 0.0028,model,0.0514 ± 0.0012,0.0344 ± 0.0006
1,machin,0.0173 ± 0.002,0.0128 ± 0.0005,new,0.0179 ± 0.0014,0.013 ± 0.0006,measur,0.0107 ± 0.0009,0.0068 ± 0.0005,learn,0.0494 ± 0.0049,0.0329 ± 0.0032,learn,0.043 ± 0.0011,0.0301 ± 0.0005
2,variabl,0.0117 ± 0.002,0.0075 ± 0.0005,cluster,0.0195 ± 0.0033,0.0152 ± 0.0016,bound,0.0154 ± 0.0012,0.0126 ± 0.0008,deep,0.0219 ± 0.004,0.0075 ± 0.0015,bound,0.0157 ± 0.0009,0.0059 ± 0.0003
3,visual,0.0095 ± 0.0021,0.0054 ± 0.0005,algorithm,0.0423 ± 0.0026,0.0385 ± 0.0013,target,0.0078 ± 0.0009,0.005 ± 0.0005,label,0.0212 ± 0.0061,0.0071 ± 0.0019,neural,0.0248 ± 0.0009,0.0151 ± 0.0004
4,imag,0.0183 ± 0.0032,0.0144 ± 0.0009,imag,0.0157 ± 0.0025,0.012 ± 0.0011,domain,0.0106 ± 0.001,0.0079 ± 0.0006,optim,0.0247 ± 0.0038,0.0112 ± 0.0021,optim,0.0218 ± 0.0009,0.0124 ± 0.0004
5,deep,0.0185 ± 0.0025,0.0147 ± 0.0007,kernel,0.0133 ± 0.0026,0.0097 ± 0.0013,languag,0.01 ± 0.0009,0.0077 ± 0.0006,classif,0.0216 ± 0.004,0.009 ± 0.002,gradient,0.0131 ± 0.0008,0.004 ± 0.0003
6,forest,0.0049 ± 0.0021,0.0013 ± 0.0003,number,0.015 ± 0.0014,0.0115 ± 0.0006,stochast,0.0114 ± 0.001,0.0094 ± 0.0006,sampl,0.0176 ± 0.0035,0.0063 ± 0.0018,estim,0.0153 ± 0.0009,0.0064 ± 0.0003
7,net,0.006 ± 0.0021,0.0025 ± 0.0004,region,0.0054 ± 0.0014,0.0023 ± 0.0004,invari,0.004 ± 0.0008,0.0021 ± 0.0004,function,0.023 ± 0.0046,0.012 ± 0.0023,sampl,0.0163 ± 0.0009,0.0079 ± 0.0004
8,corrupt,0.0045 ± 0.0019,0.0012 ± 0.0003,solv,0.0106 ± 0.0012,0.0076 ± 0.0006,constraint,0.0067 ± 0.0009,0.0048 ± 0.0004,convolut,0.0155 ± 0.004,0.0046 ± 0.002,train,0.0247 ± 0.0009,0.0165 ± 0.0004
9,autoencod,0.0075 ± 0.0024,0.0041 ± 0.0005,rnn,0.0057 ± 0.0017,0.0028 ± 0.0005,phrase,0.0029 ± 0.0007,0.0011 ± 0.0002,problem,0.0277 ± 0.0034,0.0177 ± 0.0022,convex,0.0099 ± 0.0008,0.0018 ± 0.0002
10,group,0.0075 ± 0.0019,0.0043 ± 0.0005,detector,0.0037 ± 0.0015,0.0009 ± 0.0003,sourc,0.0057 ± 0.0007,0.0039 ± 0.0004,network,0.0284 ± 0.004,0.019 ± 0.0027,task,0.0247 ± 0.0009,0.0169 ± 0.0004
11,reconstruct,0.0059 ± 0.0021,0.0027 ± 0.0003,work,0.0138 ± 0.0011,0.011 ± 0.0006,rbm,0.003 ± 0.0008,0.0012 ± 0.0004,effici,0.0175 ± 0.0026,0.0083 ± 0.0015,regret,0.0098 ± 0.0009,0.0021 ± 0.0002
12,cluster,0.0183 ± 0.0037,0.0152 ± 0.0012,machin,0.0141 ± 0.0013,0.0114 ± 0.0007,represent,0.0146 ± 0.001,0.0129 ± 0.0007,regular,0.0125 ± 0.004,0.0036 ± 0.0011,function,0.0194 ± 0.0009,0.0117 ± 0.0004
13,dataset,0.0176 ± 0.0017,0.0147 ± 0.0006,lowrank,0.005 ± 0.0012,0.0023 ± 0.0004,embed,0.0086 ± 0.001,0.0068 ± 0.0007,matrix,0.015 ± 0.0049,0.0063 ± 0.0018,algorithm,0.0325 ± 0.001,0.025 ± 0.0005
14,recoveri,0.0046 ± 0.0016,0.0017 ± 0.0003,adapt,0.0104 ± 0.0016,0.0077 ± 0.0007,sentenc,0.0053 ± 0.0008,0.0036 ± 0.0005,kernel,0.013 ± 0.005,0.0044 ± 0.0027,approxim,0.0142 ± 0.0008,0.0068 ± 0.0003
15,svm,0.0081 ± 0.0021,0.0052 ± 0.0006,ensembl,0.0054 ± 0.0016,0.0027 ± 0.0006,cca,0.0023 ± 0.0009,0.0006 ± 0.0002,recommend,0.0088 ± 0.0037,0.0003 ± 0.0003,stochast,0.0117 ± 0.0007,0.0044 ± 0.0002
16,affin,0.0038 ± 0.002,0.001 ± 0.0003,rbm,0.0043 ± 0.0018,0.0017 ± 0.0007,convex,0.009 ± 0.0009,0.0073 ± 0.0006,comput,0.0177 ± 0.0026,0.0095 ± 0.0018,matrix,0.0109 ± 0.0009,0.0038 ± 0.0003
17,prior,0.0079 ± 0.0019,0.0051 ± 0.0005,learn,0.0438 ± 0.0024,0.0412 ± 0.0013,previou,0.0074 ± 0.0005,0.0057 ± 0.0004,guarante,0.0093 ± 0.0023,0.0011 ± 0.0005,distribut,0.0167 ± 0.0008,0.01 ± 0.0003
18,convolut,0.0116 ± 0.0022,0.0089 ± 0.0006,music,0.0047 ± 0.0016,0.0022 ± 0.0006,regret,0.0087 ± 0.001,0.0071 ± 0.0007,number,0.0143 ± 0.0025,0.0061 ± 0.0013,gener,0.0249 ± 0.0008,0.0183 ± 0.0004
19,data,0.0327 ± 0.0026,0.0299 ± 0.0009,scale,0.0073 ± 0.0012,0.0048 ± 0.0005,averag,0.0051 ± 0.0006,0.0035 ± 0.0004,bound,0.0143 ± 0.0041,0.0062 ± 0.002,method,0.0295 ± 0.0009,0.023 ± 0.0004
20,spectral,0.0061 ± 0.0018,0.0034 ± 0.0004,user,0.0093 ± 0.0018,0.0068 ± 0.0008,annot,0.004 ± 0.0007,0.0024 ± 0.0003,robust,0.0133 ± 0.0037,0.0052 ± 0.0012,latent,0.0097 ± 0.0008,0.0032 ± 0.0002
21,better,0.0098 ± 0.0012,0.007 ± 0.0004,spectral,0.0054 ± 0.0013,0.0029 ± 0.0005,attent,0.0056 ± 0.0007,0.004 ± 0.0004,layer,0.0104 ± 0.0035,0.0025 ± 0.0011,infer,0.0142 ± 0.0008,0.0079 ± 0.0004
22,node,0.0074 ± 0.0022,0.0047 ± 0.0005,onlin,0.0104 ± 0.0017,0.008 ± 0.0008,point,0.0089 ± 0.0008,0.0074 ± 0.0005,propos,0.0271 ± 0.0026,0.0201 ± 0.002,stateoftheart,0.0132 ± 0.0005,0.0071 ± 0.0002
23,auction,0.0033 ± 0.0021,0.0006 ± 0.0002,convex,0.0102 ± 0.0018,0.0078 ± 0.0008,boltzmann,0.0028 ± 0.0006,0.0013 ± 0.0002,time,0.0162 ± 0.0032,0.0092 ± 0.0016,converg,0.01 ± 0.0007,0.004 ± 0.0002
24,similar,0.0111 ± 0.0019,0.0085 ± 0.0005,compress,0.004 ± 0.0013,0.0017 ± 0.0004,challeng,0.0082 ± 0.0006,0.0067 ± 0.0004,grid,0.0078 ± 0.0048,0.0009 ± 0.0006,linear,0.0112 ± 0.0007,0.0053 ± 0.0002
25,adaboost,0.0031 ± 0.0016,0.0005 ± 0.0002,train,0.0205 ± 0.0017,0.0183 ± 0.001,hand,0.0033 ± 0.0005,0.0017 ± 0.0002,cnn,0.0086 ± 0.0049,0.0018 ± 0.0009,guarante,0.0087 ± 0.0006,0.0029 ± 0.0002
26,confid,0.004 ± 0.0018,0.0015 ± 0.0003,instanc,0.0069 ± 0.0013,0.0046 ± 0.0006,composit,0.0037 ± 0.0006,0.0022 ± 0.0003,svm,0.0088 ± 0.0038,0.002 ± 0.001,variat,0.0085 ± 0.0007,0.0027 ± 0.0002
27,refer,0.0052 ± 0.0015,0.0027 ± 0.0003,depth,0.0032 ± 0.0013,0.001 ± 0.0003,prior,0.0061 ± 0.0007,0.0046 ± 0.0004,space,0.0149 ± 0.0037,0.0082 ± 0.0017,theoret,0.0097 ± 0.0006,0.0041 ± 0.0002
28,exact,0.0056 ± 0.0015,0.0031 ± 0.0003,decomposit,0.0048 ± 0.0013,0.0025 ± 0.0004,transfer,0.005 ± 0.0008,0.0035 ± 0.0005,bandit,0.009 ± 0.0033,0.0024 ± 0.0014,loss,0.0097 ± 0.0008,0.0041 ± 0.0003
29,stochast,0.0121 ± 0.002,0.0096 ± 0.0006,item,0.0055 ± 0.0017,0.0033 ± 0.0007,articl,0.0032 ± 0.0005,0.0017 ± 0.0003,graph,0.0145 ± 0.0046,0.0078 ± 0.0025,demonstr,0.0142 ± 0.0005,0.0088 ± 0.0002
30,studi,0.0125 ± 0.0014,0.01 ± 0.0004,audio,0.0037 ± 0.0014,0.0015 ± 0.0005,question,0.0061 ± 0.0008,0.0046 ± 0.0005,multiclass,0.0071 ± 0.003,0.0005 ± 0.0005,empir,0.01 ± 0.0005,0.0047 ± 0.0002
31,combin,0.0098 ± 0.0014,0.0074 ± 0.0004,random,0.0103 ± 0.0016,0.0081 ± 0.0008,metric,0.0067 ± 0.001,0.0053 ± 0.0006,veri,0.0098 ± 0.0024,0.0032 ± 0.0009,spars,0.0086 ± 0.0007,0.0033 ± 0.0002
32,kmean,0.0052 ± 0.0022,0.0028 ± 0.0005,unsupervis,0.007 ± 0.0016,0.0048 ± 0.0006,structur,0.015 ± 0.0011,0.0136 ± 0.0007,norm,0.0065 ± 0.0026,1e-04 ± 1e-04,regular,0.0094 ± 0.0007,0.0042 ± 0.0003
33,best,0.0078 ± 0.0011,0.0055 ± 0.0003,rate,0.0109 ± 0.0019,0.0088 ± 0.0008,belief,0.0033 ± 0.0007,0.0019 ± 0.0003,kernelbas,0.0064 ± 0.0033,0.0 ± 0.0,improv,0.0164 ± 0.0006,0.0113 ± 0.0003
34,mani,0.0119 ± 0.0012,0.0096 ± 0.0004,addit,0.0078 ± 0.0012,0.0057 ± 0.0004,signific,0.0056 ± 0.0005,0.0042 ± 0.0003,class,0.0111 ± 0.0027,0.0047 ± 0.0014,simpl,0.0108 ± 0.0005,0.0058 ± 0.0002
35,spatial,0.0047 ± 0.0015,0.0024 ± 0.0004,manipul,0.0027 ± 0.0011,0.0007 ± 0.0002,noisi,0.004 ± 0.0005,0.0026 ± 0.0003,adapt,0.0121 ± 0.0035,0.0058 ± 0.0016,achiev,0.0139 ± 0.0005,0.0089 ± 0.0002
36,convex,0.0104 ± 0.0023,0.0082 ± 0.0006,semisupervis,0.0047 ± 0.0015,0.0027 ± 0.0005,provid,0.014 ± 0.0007,0.0127 ± 0.0005,constant,0.0065 ± 0.0021,0.0002 ± 0.0002,kernel,0.0092 ± 0.001,0.0042 ± 0.0004
37,composit,0.005 ± 0.0014,0.0027 ± 0.0004,real,0.007 ± 0.0009,0.005 ± 0.0004,hierarch,0.0046 ± 0.0007,0.0032 ± 0.0004,exist,0.0164 ± 0.0027,0.0102 ± 0.0016,word,0.0195 ± 0.0011,0.0149 ± 0.0006
38,research,0.0078 ± 0.0014,0.0056 ± 0.0004,reinforc,0.0073 ± 0.0014,0.0053 ± 0.0006,extract,0.0067 ± 0.0007,0.0054 ± 0.0004,onlin,0.0106 ± 0.0027,0.0045 ± 0.0015,sequenc,0.0114 ± 0.0008,0.0069 ± 0.0004
39,gradient,0.0113 ± 0.002,0.009 ± 0.0006,partit,0.0039 ± 0.0013,0.0019 ± 0.0004,energi,0.0037 ± 0.0008,0.0024 ± 0.0003,consid,0.0142 ± 0.0023,0.008 ± 0.0015,predict,0.0163 ± 0.0008,0.012 ± 0.0004
40,onlin,0.0111 ± 0.0023,0.0089 ± 0.0007,ssc,0.0021 ± 0.0015,0.0002 ± 0.0002,repres,0.006 ± 0.0005,0.0047 ± 0.0003,continu,0.0094 ± 0.0033,0.0034 ± 0.0012,bandit,0.0058 ± 0.0007,0.0016 ± 0.0002
41,subspac,0.0054 ± 0.0026,0.0032 ± 0.0005,net,0.0041 ± 0.0012,0.0022 ± 0.0005,drift,0.0021 ± 0.0007,0.0008 ± 0.0003,corrupt,0.0068 ± 0.0033,0.0008 ± 0.0006,depend,0.0105 ± 0.0006,0.0063 ± 0.0003
42,memori,0.0081 ± 0.0019,0.006 ± 0.0005,select,0.0118 ± 0.0017,0.0099 ± 0.0008,translat,0.0053 ± 0.0008,0.004 ± 0.0006,point,0.0124 ± 0.0038,0.0065 ± 0.0016,network,0.0298 ± 0.0011,0.0256 ± 0.0006
43,realworld,0.0069 ± 0.0011,0.0047 ± 0.0003,mechan,0.0053 ± 0.0012,0.0034 ± 0.0005,unknown,0.0046 ± 0.0006,0.0034 ± 0.0004,case,0.0114 ± 0.0022,0.0055 ± 0.0012,translat,0.0123 ± 0.001,0.0082 ± 0.0005
44,power,0.0073 ± 0.0012,0.0051 ± 0.0004,analysi,0.0117 ± 0.0012,0.0098 ± 0.0006,gap,0.003 ± 0.0005,0.0018 ± 0.0002,potenti,0.0087 ± 0.0021,0.0029 ± 0.0009,sentenc,0.011 ± 0.0009,0.0069 ± 0.0004
45,motif,0.0029 ± 0.0022,0.0008 ± 0.0003,confid,0.0031 ± 0.0012,0.0012 ± 0.0003,opinion,0.0019 ± 0.0006,0.0007 ± 0.0002,rnn,0.008 ± 0.0042,0.0022 ± 0.001,problem,0.0257 ± 0.0008,0.0216 ± 0.0004
46,signal,0.0068 ± 0.0017,0.0046 ± 0.0005,bandit,0.007 ± 0.0018,0.0051 ± 0.0008,dirichlet,0.002 ± 0.0005,0.0008 ± 0.0002,new,0.0163 ± 0.0022,0.0106 ± 0.0015,pars,0.0069 ± 0.0007,0.0029 ± 0.0003
47,text,0.0087 ± 0.0023,0.0065 ± 0.0005,regret,0.0102 ± 0.002,0.0083 ± 0.001,syntact,0.0021 ± 0.0004,0.0008 ± 0.0002,review,0.0076 ± 0.0049,0.0019 ± 0.0009,paramet,0.0113 ± 0.0006,0.0073 ± 0.0003
48,particularli,0.0035 ± 0.0009,0.0014 ± 0.0002,matric,0.0055 ± 0.0013,0.0037 ± 0.0006,recoveri,0.0027 ± 0.0005,0.0014 ± 0.0003,predict,0.0164 ± 0.0035,0.0108 ± 0.0022,nonconvex,0.0048 ± 0.0006,0.0008 ± 1e-04
49,librari,0.003 ± 0.0015,0.0009 ± 0.0002,dcop,0.0019 ± 0.0014,1e-04 ± 1e-04,mdp,0.0023 ± 0.0007,0.0011 ± 0.0003,data,0.0296 ± 0.0038,0.024 ± 0.0026,outperform,0.0098 ± 0.0004,0.0059 ± 0.0002
\end{filecontents*}
\csvautotabular{data/abstract_words_acc.csv}
		}
		\caption{Most important predictors for acceptance in decreasing order of importance: abstract} \label{abstract_words_acc}
    \end{table*}
    \end{landscape}
    \clearpage
}

\afterpage{%
    \clearpage
    \thispagestyle{empty}
    \begin{landscape}
    \begin{table*}
        \centering 
        \resizebox{\columnwidth}{170pt}{
		\begin{filecontents*}{data/introduction_words_acc.csv}
,AI Stem,AI Score accepted,AI Score rejected,CL Stem,CL Score accepted,CL Score rejected,LG Stem,LG Score accepted,LG Score rejected,ICLR Stem,ICLR Score accepted,ICLR Score rejected,All Stem,All Score accepted,All Score rejected
0,layer,0.0174 ± 0.003,0.012 ± 0.0008,learn,0.0546 ± 0.0025,0.046 ± 0.0014,model,0.0496 ± 0.0015,0.0437 ± 0.0011,network,0.0373 ± 0.0048,0.0132 ± 0.0018,model,0.0578 ± 0.0012,0.0405 ± 0.0006
1,word,0.0208 ± 0.0034,0.0157 ± 0.0011,label,0.0242 ± 0.0025,0.0175 ± 0.0014,word,0.0178 ± 0.0015,0.0122 ± 0.0009,algorithm,0.0406 ± 0.0041,0.0198 ± 0.002,learn,0.0484 ± 0.001,0.0337 ± 0.0006
2,vector,0.0188 ± 0.0025,0.0142 ± 0.0007,train,0.0284 ± 0.0019,0.0224 ± 0.001,languag,0.0141 ± 0.001,0.0097 ± 0.0006,learn,0.0508 ± 0.0041,0.0302 ± 0.0028,estim,0.0188 ± 0.0008,0.0073 ± 0.0003
3,label,0.0233 ± 0.0032,0.0189 ± 0.0011,class,0.0177 ± 0.0017,0.0123 ± 0.0008,object,0.0176 ± 0.001,0.0141 ± 0.0007,matrix,0.0226 ± 0.0051,0.0048 ± 0.0014,optim,0.024 ± 0.0008,0.0133 ± 0.0004
4,activ,0.0123 ± 0.0022,0.0082 ± 0.0006,classifi,0.017 ± 0.0019,0.0117 ± 0.0009,sentenc,0.008 ± 0.0009,0.0049 ± 0.0005,kernel,0.0209 ± 0.0065,0.0038 ± 0.0026,function,0.0257 ± 0.0008,0.0152 ± 0.0004
5,speech,0.0093 ± 0.002,0.0052 ± 0.0005,imag,0.0225 ± 0.0025,0.0174 ± 0.0013,document,0.0094 ± 0.001,0.0065 ± 0.0007,label,0.0253 ± 0.0059,0.0088 ± 0.0023,bound,0.0177 ± 0.0008,0.0076 ± 0.0004
6,bit,0.0061 ± 0.0018,0.0022 ± 0.0004,classif,0.018 ± 0.0016,0.0133 ± 0.0008,annot,0.0061 ± 0.0007,0.0033 ± 0.0004,deep,0.0219 ± 0.0038,0.0056 ± 0.0011,gradient,0.0153 ± 0.0008,0.0056 ± 0.0003
7,lstm,0.0078 ± 0.0025,0.004 ± 0.0005,kernel,0.0196 ± 0.0032,0.0152 ± 0.0017,arm,0.0072 ± 0.0012,0.0045 ± 0.0008,adversari,0.0159 ± 0.0057,0.0013 ± 0.001,sampl,0.02 ± 0.0008,0.0104 ± 0.0004
8,neural,0.0196 ± 0.0022,0.0158 ± 0.0007,margin,0.0092 ± 0.0016,0.005 ± 0.0006,event,0.0067 ± 0.0011,0.004 ± 0.0005,function,0.0274 ± 0.0045,0.0137 ± 0.0021,matrix,0.0157 ± 0.0009,0.0063 ± 0.0003
9,predict,0.0217 ± 0.0026,0.018 ± 0.0008,represent,0.0189 ± 0.0018,0.0149 ± 0.0009,entiti,0.0066 ± 0.0009,0.004 ± 0.0006,imag,0.0246 ± 0.0049,0.0114 ± 0.003,distribut,0.0219 ± 0.0008,0.0125 ± 0.0004
10,represent,0.02 ± 0.0022,0.0163 ± 0.0008,variat,0.0087 ± 0.0014,0.0051 ± 0.0006,graph,0.0131 ± 0.0013,0.0106 ± 0.0009,bound,0.0195 ± 0.0037,0.0068 ± 0.0019,algorithm,0.035 ± 0.0009,0.0257 ± 0.0005
11,answer,0.0098 ± 0.0022,0.0061 ± 0.0006,face,0.0067 ± 0.0013,0.0031 ± 0.0005,vector,0.0155 ± 0.0009,0.0131 ± 0.0006,train,0.0277 ± 0.0036,0.015 ± 0.0017,train,0.0294 ± 0.0008,0.0211 ± 0.0005
12,bandit,0.0096 ± 0.0027,0.006 ± 0.0007,converg,0.0118 ± 0.0013,0.0083 ± 0.0007,embed,0.0108 ± 0.0011,0.0084 ± 0.0008,convolut,0.0139 ± 0.0037,0.0017 ± 0.0006,convex,0.0108 ± 0.0007,0.0026 ± 0.0002
13,video,0.0077 ± 0.002,0.0042 ± 0.0006,updat,0.0091 ± 0.0012,0.0056 ± 0.0005,caption,0.0037 ± 0.0008,0.0015 ± 0.0003,optim,0.022 ± 0.0032,0.0101 ± 0.0015,approxim,0.016 ± 0.0007,0.0078 ± 0.0003
14,scale,0.0106 ± 0.0017,0.0072 ± 0.0004,ensembl,0.0059 ± 0.0018,0.0025 ± 0.0005,path,0.0049 ± 0.0007,0.0026 ± 0.0004,sampl,0.0189 ± 0.0032,0.0073 ± 0.0017,neural,0.0238 ± 0.0008,0.0156 ± 0.0004
15,unlabel,0.0072 ± 0.0016,0.0037 ± 0.0004,predict,0.02 ± 0.0019,0.0166 ± 0.001,label,0.0204 ± 0.0015,0.0182 ± 0.0011,neural,0.0219 ± 0.0035,0.0107 ± 0.002,loss,0.014 ± 0.0008,0.006 ± 0.0004
16,cnn,0.0111 ± 0.0025,0.0078 ± 0.0008,multiclass,0.006 ± 0.0013,0.0027 ± 0.0005,translat,0.0075 ± 0.0009,0.0054 ± 0.0005,layer,0.0154 ± 0.003,0.0043 ± 0.0011,regret,0.011 ± 0.001,0.0031 ± 0.0003
17,perform,0.0204 ± 0.0012,0.0171 ± 0.0004,regular,0.0131 ± 0.0015,0.0098 ± 0.0009,submodular,0.0045 ± 0.0011,0.0025 ± 0.0007,bandit,0.0137 ± 0.0048,0.0029 ± 0.0015,stochast,0.0114 ± 0.0006,0.0042 ± 0.0002
18,transform,0.0097 ± 0.0019,0.0064 ± 0.0005,elast,0.004 ± 0.0017,0.0007 ± 0.0003,nlp,0.0038 ± 0.0004,0.0019 ± 0.0003,arm,0.0118 ± 0.0047,0.0013 ± 0.001,infer,0.0159 ± 0.0008,0.0088 ± 0.0004
19,learn,0.0531 ± 0.0031,0.05 ± 0.0011,data,0.0354 ± 0.0019,0.0323 ± 0.0011,attent,0.0069 ± 0.0009,0.005 ± 0.0004,regular,0.013 ± 0.0043,0.0029 ± 0.0006,kernel,0.0128 ± 0.0012,0.0057 ± 0.0005
20,imag,0.0239 ± 0.0033,0.0208 ± 0.0012,adr,0.003 ± 0.0018,0.0 ± 0.0,arab,0.002 ± 0.0007,0.0002 ± 0.0002,regret,0.0149 ± 0.005,0.005 ± 0.002,paramet,0.0153 ± 0.0006,0.0086 ± 0.0003
21,speaker,0.0045 ± 0.0021,0.0014 ± 0.0004,sequenc,0.0114 ± 0.0018,0.0085 ± 0.0007,descriptor,0.0026 ± 0.0007,0.001 ± 0.0003,graph,0.0198 ± 0.0056,0.0104 ± 0.0025,latent,0.0112 ± 0.0008,0.0047 ± 0.0003
22,regret,0.0135 ± 0.0033,0.0104 ± 0.001,robust,0.0079 ± 0.0013,0.005 ± 0.0004,probabilist,0.0059 ± 0.0005,0.0043 ± 0.0003,architectur,0.0134 ± 0.003,0.0041 ± 0.0011,linear,0.0126 ± 0.0005,0.0061 ± 0.0002
23,memori,0.0098 ± 0.0021,0.0067 ± 0.0006,learner,0.0086 ± 0.0016,0.0058 ± 0.0007,imag,0.0236 ± 0.0016,0.022 ± 0.0013,paramet,0.0153 ± 0.0022,0.0061 ± 0.0012,regular,0.0114 ± 0.0007,0.005 ± 0.0003
24,recognit,0.0119 ± 0.0017,0.0088 ± 0.0006,loss,0.0158 ± 0.002,0.0129 ± 0.0011,local,0.0105 ± 0.0008,0.0089 ± 0.0005,reward,0.0175 ± 0.0049,0.0086 ± 0.0034,method,0.0277 ± 0.0007,0.0215 ± 0.0004
25,cluster,0.0218 ± 0.0046,0.0188 ± 0.0015,svm,0.0092 ± 0.0019,0.0064 ± 0.0008,regret,0.0114 ± 0.0013,0.0098 ± 0.001,classifi,0.0159 ± 0.0032,0.0073 ± 0.0014,gener,0.0265 ± 0.0007,0.0203 ± 0.0003
26,caption,0.0047 ± 0.0022,0.0017 ± 0.0004,estim,0.0188 ± 0.0017,0.016 ± 0.001,text,0.009 ± 0.0008,0.0075 ± 0.0006,norm,0.0109 ± 0.0027,0.0023 ± 0.0011,vector,0.0166 ± 0.0008,0.0105 ± 0.0004
27,distanc,0.0096 ± 0.0021,0.0066 ± 0.0005,nonlinear,0.0079 ± 0.001,0.0051 ± 0.0005,supervis,0.0096 ± 0.0007,0.008 ± 0.0005,gener,0.0265 ± 0.0029,0.018 ± 0.0016,converg,0.0104 ± 0.0006,0.0043 ± 0.0002
28,bengio,0.0063 ± 0.0017,0.0033 ± 0.0003,dropout,0.005 ± 0.0019,0.0023 ± 0.0007,causal,0.0038 ± 0.0009,0.0022 ± 0.0005,svm,0.0109 ± 0.0045,0.0026 ± 0.0011,spars,0.01 ± 0.0007,0.004 ± 0.0003
29,topic,0.0098 ± 0.0028,0.0069 ± 0.0008,cover,0.0048 ± 0.001,0.0022 ± 0.0002,market,0.003 ± 0.0007,0.0015 ± 0.0003,set,0.0256 ± 0.0026,0.0175 ± 0.0015,predict,0.0192 ± 0.0008,0.0134 ± 0.0004
30,convex,0.0127 ± 0.0021,0.0099 ± 0.0007,visual,0.0082 ± 0.0013,0.0055 ± 0.0006,structur,0.0159 ± 0.0009,0.0144 ± 0.0006,stochast,0.0116 ± 0.0025,0.0037 ± 0.001,bandit,0.0074 ± 0.0008,0.0018 ± 0.0002
31,translat,0.0088 ± 0.0023,0.0059 ± 0.0006,dictionari,0.0061 ± 0.0015,0.0034 ± 0.0007,mixtur,0.0054 ± 0.0008,0.0039 ± 0.0005,confid,0.0086 ± 0.0041,0.0009 ± 0.0004,network,0.0328 ± 0.0011,0.0272 ± 0.0006
32,weight,0.0134 ± 0.0018,0.0106 ± 0.0006,exampl,0.0164 ± 0.001,0.0138 ± 0.0005,covari,0.0049 ± 0.0007,0.0034 ± 0.0004,weight,0.0136 ± 0.0026,0.0059 ± 0.0015,variat,0.0087 ± 0.0006,0.0034 ± 0.0002
33,microphon,0.0028 ± 0.0019,0.0 ± 0.0,function,0.0273 ± 0.0019,0.0247 ± 0.0011,corpora,0.0026 ± 0.0004,0.0011 ± 0.0002,minim,0.0117 ± 0.0019,0.004 ± 0.0009,guarante,0.0086 ± 0.0004,0.0035 ± 0.0002
34,dataset,0.0162 ± 0.0016,0.0134 ± 0.0006,detector,0.0042 ± 0.0012,0.0017 ± 0.0004,phrase,0.0035 ± 0.0006,0.002 ± 0.0004,convex,0.0099 ± 0.0025,0.0028 ± 0.0009,polici,0.0123 ± 0.001,0.0073 ± 0.0005
35,theano,0.0032 ± 0.0021,0.0004 ± 1e-04,infer,0.0129 ± 0.0016,0.0104 ± 0.0008,semant,0.0083 ± 0.0007,0.0069 ± 0.0005,perturb,0.0074 ± 0.0032,0.0003 ± 0.0002,task,0.0244 ± 0.0008,0.0194 ± 0.0004
36,sourc,0.0085 ± 0.0019,0.0059 ± 0.0005,tempor,0.0061 ± 0.0012,0.0037 ± 0.0005,spl,0.0014 ± 0.0008,0.0 ± 0.0,predict,0.0186 ± 0.0037,0.0116 ± 0.002,norm,0.0072 ± 0.0007,0.0022 ± 0.0002
37,decod,0.0053 ± 0.0015,0.0027 ± 0.0004,assign,0.0066 ± 0.001,0.0042 ± 0.0004,depend,0.011 ± 0.0006,0.0097 ± 0.0005,flow,0.0079 ± 0.0045,0.0012 ± 0.0004,label,0.0172 ± 0.001,0.0123 ± 0.0005
38,dropout,0.0056 ± 0.0026,0.0031 ± 0.0006,section,0.0275 ± 0.0014,0.025 ± 0.0008,diffus,0.0023 ± 0.0006,0.0009 ± 0.0003,hash,0.007 ± 0.0046,0.0004 ± 0.0003,random,0.0108 ± 0.0005,0.0059 ± 0.0002
39,atp,0.0029 ± 0.0018,0.0003 ± 0.0002,multipl,0.0092 ± 0.0008,0.0068 ± 0.0004,interact,0.0066 ± 0.0006,0.0052 ± 0.0004,data,0.0315 ± 0.0035,0.0249 ± 0.0027,posterior,0.0064 ± 0.0006,0.0016 ± 0.0002
40,phrase,0.0055 ± 0.0018,0.003 ± 0.0005,modal,0.0034 ± 0.0011,0.0011 ± 0.0004,question,0.009 ± 0.0008,0.0077 ± 0.0006,featur,0.0236 ± 0.0036,0.017 ± 0.0027,sequenc,0.0144 ± 0.0008,0.0095 ± 0.0004
41,dae,0.0029 ± 0.0018,0.0004 ± 0.0002,approxim,0.0157 ± 0.0014,0.0135 ± 0.0008,privaci,0.0048 ± 0.001,0.0034 ± 0.0007,batch,0.0076 ± 0.0032,0.0011 ± 0.0006,sentenc,0.0164 ± 0.001,0.0116 ± 0.0005
42,gpu,0.0053 ± 0.0018,0.0028 ± 0.0004,spike,0.0028 ± 0.0011,0.0005 ± 0.0002,inform,0.0156 ± 0.0006,0.0143 ± 0.0005,loss,0.0134 ± 0.0031,0.007 ± 0.0023,nmt,0.0075 ± 0.001,0.0028 ± 0.0004
43,path,0.0059 ± 0.0017,0.0034 ± 0.0005,channel,0.0032 ± 0.0011,0.001 ± 0.0002,price,0.0033 ± 0.0007,0.002 ± 0.0003,approxim,0.0134 ± 0.0025,0.007 ± 0.0014,pars,0.0091 ± 0.0008,0.0044 ± 0.0003
44,unsupervis,0.0092 ± 0.0016,0.0068 ± 0.0005,cnl,0.0023 ± 0.0016,0.0 ± 0.0,smooth,0.0053 ± 0.0006,0.0041 ± 0.0004,iter,0.0082 ± 0.002,0.0019 ± 0.0006,empir,0.0084 ± 0.0003,0.0038 ± 1e-04
45,feedback,0.0066 ± 0.002,0.0042 ± 0.0006,forecast,0.0029 ± 0.0014,0.0008 ± 0.0002,nmt,0.0017 ± 0.0006,0.0004 ± 0.0002,crowdsourc,0.0067 ± 0.004,0.0006 ± 0.0003,assumpt,0.0091 ± 0.0004,0.0045 ± 0.0002
46,autoencod,0.0071 ± 0.0021,0.0047 ± 0.0006,scene,0.0044 ± 0.0015,0.0023 ± 0.0005,context,0.0081 ± 0.0006,0.0068 ± 0.0004,permut,0.006 ± 0.0024,0.0 ± 0.0,reward,0.0099 ± 0.0008,0.0054 ± 0.0004
47,kernel,0.0178 ± 0.0043,0.0154 ± 0.0013,textur,0.0026 ± 0.0013,0.0004 ± 1e-04,som,0.0016 ± 0.0006,0.0004 ± 0.0002,rate,0.0128 ± 0.0031,0.0068 ± 0.0016,depend,0.0122 ± 0.0005,0.0078 ± 0.0002
48,question,0.0101 ± 0.0018,0.0077 ± 0.0006,sampl,0.0212 ± 0.0018,0.0191 ± 0.0011,link,0.004 ± 0.0006,0.0028 ± 0.0003,class,0.0127 ± 0.0026,0.0068 ± 0.0012,gaussian,0.0069 ± 0.0005,0.0025 ± 0.0002
49,request,0.0034 ± 0.0013,0.0011 ± 0.0002,framework,0.0116 ± 0.0009,0.0094 ± 0.0005,global,0.0057 ± 0.0005,0.0045 ± 0.0003,log,0.0072 ± 0.0017,0.0013 ± 0.0005,observ,0.0125 ± 0.0005,0.0082 ± 0.0003
\end{filecontents*}
\csvautotabular{data/introduction_words_acc.csv}
		}
		\caption{Most important predictors for acceptance in decreasing order of importance: introduction} \label{introduction_words_acc}
    \end{table*}
    \end{landscape}
    \clearpage
}

\afterpage{%
    \clearpage
    \thispagestyle{empty}
    \begin{landscape}
    \begin{table*}
        \centering 
        \resizebox{\columnwidth}{170pt}{
		\begin{filecontents*}{data/title_words_rej.csv}
,AI Stem,AI Score accepted,AI Score rejected,CL Stem,CL Score accepted,CL Score rejected,LG Stem,LG Score accepted,LG Score rejected,ICLR Stem,ICLR Score accepted,ICLR Score rejected,All Stem,All Score accepted,All Score rejected
0,rank,0.0011 ± 0.0011,0.0062 ± 0.0009,network,0.0223 ± 0.0032,0.0283 ± 0.0018,network,0.0305 ± 0.002,0.0361 ± 0.0016,deep,0.0163 ± 0.0066,0.0286 ± 0.007,use,0.0113 ± 0.0009,0.0204 ± 0.0007
1,learn,0.0459 ± 0.0043,0.051 ± 0.0016,bayesian,0.0063 ± 0.0019,0.0121 ± 0.0015,deep,0.0236 ± 0.0019,0.0268 ± 0.0015,hash,0.0 ± 0.0,0.0113 ± 0.0066,base,0.0035 ± 0.0006,0.0125 ± 0.0006
2,reinforc,0.0033 ± 0.0019,0.0079 ± 0.0009,gener,0.0088 ± 0.0021,0.0139 ± 0.0015,adapt,0.0078 ± 0.0012,0.0108 ± 0.0011,risk,0.0 ± 0.0,0.0113 ± 0.0057,recognit,0.0026 ± 0.0006,0.0093 ± 0.0006
3,robust,0.0009 ± 0.0009,0.005 ± 0.0008,classif,0.0075 ± 0.002,0.0115 ± 0.0014,estim,0.0062 ± 0.0011,0.0088 ± 0.001,explor,0.0 ± 0.0,0.0105 ± 0.0062,detect,0.0029 ± 0.0006,0.0096 ± 0.0006
4,distribut,0.0055 ± 0.0021,0.0094 ± 0.001,recurr,0.0027 ± 0.0012,0.0068 ± 0.0012,framework,0.0032 ± 0.0008,0.0058 ± 0.0008,neural,0.0248 ± 0.0072,0.0353 ± 0.0074,speech,0.0009 ± 0.0004,0.0065 ± 0.0005
5,process,0.0027 ± 0.0016,0.0064 ± 0.0008,imag,0.0042 ± 0.0015,0.0082 ± 0.0013,connect,0.0004 ± 0.0003,0.0029 ± 0.0007,bayesian,0.0062 ± 0.0045,0.0165 ± 0.0068,approach,0.0057 ± 0.0008,0.0112 ± 0.0006
6,matrix,0.0025 ± 0.0014,0.0062 ± 0.0009,dynam,0.0032 ± 0.0014,0.007 ± 0.0012,fast,0.0056 ± 0.0011,0.0081 ± 0.001,deriv,0.0 ± 0.0,0.01 ± 0.0059,logic,0.0016 ± 0.0004,0.0071 ± 0.0006
7,code,0.0 ± 0.0,0.0037 ± 0.0007,nonparametr,0.0007 ± 0.0007,0.0046 ± 0.0011,perform,0.0016 ± 0.0006,0.0039 ± 0.0007,continu,0.0 ± 0.0,0.0095 ± 0.0056,data,0.0071 ± 0.0009,0.0122 ± 0.0006
8,word,0.0014 ± 0.0011,0.0051 ± 0.0007,tempor,0.0 ± 0.0,0.0037 ± 0.0009,state,0.001 ± 0.0005,0.0033 ± 0.0007,multiarm,0.0 ± 0.0,0.009 ± 0.0065,classif,0.0059 ± 0.0008,0.0109 ± 0.0006
9,gaussian,0.0022 ± 0.0015,0.0058 ± 0.0009,method,0.0065 ± 0.0018,0.0101 ± 0.0013,classifi,0.0043 ± 0.001,0.0064 ± 0.0009,trace,0.0 ± 0.0,0.0086 ± 0.0062,plan,0.0019 ± 0.0005,0.0066 ± 0.0005
10,compress,0.0 ± 0.0,0.0034 ± 0.0007,word,0.0033 ± 0.0014,0.0065 ± 0.0011,recurr,0.0084 ± 0.0013,0.0105 ± 0.0012,twin,0.0 ± 0.0,0.0086 ± 0.006,intellig,0.0005 ± 0.0003,0.005 ± 0.0005
11,adversari,0.0 ± 0.0,0.0034 ± 0.0007,game,0.0 ± 0.0,0.0032 ± 0.0009,recognit,0.0058 ± 0.001,0.0079 ± 0.0009,spammer,0.0 ± 0.0,0.0084 ± 0.006,fuzzi,0.0 ± 0.0,0.0045 ± 0.0005
12,polici,0.0 ± 0.0,0.0033 ± 0.0006,rank,0.0032 ± 0.0014,0.0062 ± 0.0012,similar,0.0011 ± 0.0005,0.0032 ± 0.0007,multiview,0.0 ± 0.0,0.0083 ± 0.0048,reason,0.0017 ± 0.0005,0.0061 ± 0.0005
13,complex,0.001 ± 0.001,0.0043 ± 0.0007,tree,0.0024 ± 0.0012,0.0054 ± 0.0011,video,0.0019 ± 0.0007,0.004 ± 0.0007,gate,0.0 ± 0.0,0.0083 ± 0.0058,belief,0.0011 ± 0.0004,0.0054 ± 0.0005
14,submodular,0.0 ± 0.0,0.0032 ± 0.0007,dual,0.0 ± 0.0,0.0027 ± 0.0008,big,0.0012 ± 0.0005,0.0033 ± 0.0007,contextawar,0.0 ± 0.0,0.0082 ± 0.0048,survey,0.0002 ± 0.0002,0.0042 ± 0.0005
15,converg,0.0013 ± 0.0013,0.0045 ± 0.0007,multilabel,0.0007 ± 0.0007,0.0033 ± 0.001,analysi,0.0064 ± 0.0011,0.0085 ± 0.0009,point,0.0 ± 0.0,0.0078 ± 0.0055,probabilist,0.0038 ± 0.0007,0.0077 ± 0.0006
16,iter,0.0 ± 0.0,0.0031 ± 0.0006,train,0.0036 ± 0.0015,0.0062 ± 0.0011,imag,0.0063 ± 0.0012,0.0083 ± 0.001,optim,0.0091 ± 0.0069,0.0167 ± 0.006,inform,0.0028 ± 0.0006,0.0068 ± 0.0005
17,kernel,0.0074 ± 0.0026,0.0104 ± 0.0012,paramet,0.0022 ± 0.0013,0.0048 ± 0.0011,scene,0.0004 ± 0.0003,0.0022 ± 0.0006,net,0.0 ± 0.0,0.0075 ± 0.0053,analysi,0.0062 ± 0.0008,0.01 ± 0.0006
18,transform,0.0 ± 0.0,0.003 ± 0.0006,interact,0.0006 ± 0.0006,0.0032 ± 0.0008,time,0.0034 ± 0.0008,0.0052 ± 0.0008,languag,0.0038 ± 0.0038,0.0112 ± 0.005,robot,0.0002 ± 1e-04,0.0039 ± 0.0004
19,segment,0.0 ± 0.0,0.0029 ± 0.0006,search,0.0025 ± 0.0011,0.0051 ± 0.001,rank,0.0046 ± 0.0011,0.0063 ± 0.0009,budget,0.0 ± 0.0,0.0074 ± 0.0053,automat,0.0011 ± 0.0004,0.0048 ± 0.0004
20,net,0.0 ± 0.0,0.0029 ± 0.0006,practic,0.0 ± 0.0,0.0026 ± 0.0008,architectur,0.0023 ± 0.0007,0.004 ± 0.0007,quantiz,0.0 ± 0.0,0.0073 ± 0.0052,ontolog,0.0003 ± 0.0002,0.0038 ± 0.0004
21,problem,0.003 ± 0.0015,0.0059 ± 0.0008,data,0.0107 ± 0.0022,0.0133 ± 0.0014,tensor,0.0019 ± 0.0007,0.0036 ± 0.0007,perturb,0.0 ± 0.0,0.0073 ± 0.0052,knowledg,0.0046 ± 0.0008,0.0078 ± 0.0006
22,similar,0.0 ± 0.0,0.0027 ± 0.0006,discoveri,0.0012 ± 0.0008,0.0037 ± 0.0009,whi,0.0 ± 0.0,0.0016 ± 0.0005,markov,0.0 ± 0.0,0.0072 ± 0.0041,comput,0.0034 ± 0.0007,0.0066 ± 0.0005
23,label,0.0018 ± 0.0013,0.0045 ± 0.0008,unsupervi,0.0032 ± 0.0015,0.0057 ± 0.0011,pattern,0.0009 ± 0.0004,0.0025 ± 0.0006,pixel,0.0 ± 0.0,0.0071 ± 0.0071,deci,0.0024 ± 0.0006,0.0055 ± 0.0005
24,solv,0.0 ± 0.0,0.0027 ± 0.0006,annot,0.0 ± 0.0,0.0024 ± 0.0008,person,0.0011 ± 0.0005,0.0027 ± 0.0006,transfer,0.0 ± 0.0,0.007 ± 0.0049,techniqu,0.0007 ± 0.0003,0.0038 ± 0.0004
25,autoencod,0.0023 ± 0.0016,0.0049 ± 0.0008,variabl,0.0007 ± 0.0007,0.0031 ± 0.0008,varianc,0.0003 ± 0.0003,0.0019 ± 0.0005,batch,0.0 ± 0.0,0.0069 ± 0.0053,web,0.0004 ± 0.0002,0.0034 ± 0.0004
26,activ,0.0033 ± 0.0017,0.0059 ± 0.0008,multitask,0.0017 ± 0.0012,0.0041 ± 0.001,trace,0.0 ± 0.0,0.0016 ± 0.0006,multicor,0.0 ± 0.0,0.0069 ± 0.0069,studi,0.0014 ± 0.0004,0.0043 ± 0.0004
27,maxim,0.0 ± 0.0,0.0026 ± 0.0006,event,0.0017 ± 0.001,0.0041 ± 0.0009,new,0.0021 ± 0.0007,0.0037 ± 0.0007,viewpoint,0.0 ± 0.0,0.0067 ± 0.0067,evalu,0.0032 ± 0.0006,0.006 ± 0.0005
28,feedback,0.0008 ± 0.0008,0.0034 ± 0.0007,construct,0.0 ± 0.0,0.0023 ± 0.0007,distanc,0.0007 ± 0.0004,0.0023 ± 0.0006,tree,0.0 ± 0.0,0.0067 ± 0.0048,framework,0.0026 ± 0.0006,0.0053 ± 0.0005
29,convex,0.0034 ± 0.002,0.006 ± 0.0009,mixtur,0.0023 ± 0.0014,0.0045 ± 0.0011,action,0.0008 ± 0.0005,0.0023 ± 0.0006,financ,0.0 ± 0.0,0.0066 ± 0.0066,review,0.0005 ± 0.0003,0.0032 ± 0.0004
30,interact,0.0 ± 0.0,0.0025 ± 0.0005,variat,0.0033 ± 0.0015,0.0054 ± 0.0011,repr,0.0077 ± 0.0012,0.0093 ± 0.001,nonneg,0.0033 ± 0.0033,0.0099 ± 0.0058,autom,0.0003 ± 0.0002,0.0029 ± 0.0004
31,empir,0.0 ± 0.0,0.0024 ± 0.0006,plan,0.0019 ± 0.0011,0.004 ± 0.0009,autoencod,0.0035 ± 0.001,0.005 ± 0.0008,lstm,0.0 ± 0.0,0.0066 ± 0.0047,set,0.0022 ± 0.0005,0.0048 ± 0.0004
32,guarant,0.0 ± 0.0,0.0023 ± 0.0006,error,0.0007 ± 0.0007,0.0028 ± 0.0008,effect,0.0008 ± 0.0004,0.0024 ± 0.0005,opposit,0.0 ± 0.0,0.0065 ± 0.0065,artifici,0.0004 ± 0.0002,0.003 ± 0.0004
33,bandit,0.0057 ± 0.0023,0.008 ± 0.001,categor,0.0 ± 0.0,0.0021 ± 0.0007,convolut,0.0109 ± 0.0015,0.0124 ± 0.0012,minim,0.0 ± 0.0,0.0064 ± 0.0045,perform,0.0009 ± 0.0003,0.0033 ± 0.0004
34,algorithm,0.0132 ± 0.0029,0.0155 ± 0.0011,hybrid,0.0006 ± 0.0006,0.0027 ± 0.0008,solut,0.0004 ± 0.0003,0.0019 ± 0.0005,clinic,0.0 ± 0.0,0.0063 ± 0.0046,uncertainti,0.0007 ± 0.0003,0.0031 ± 0.0004
35,constrain,0.0 ± 0.0,0.0023 ± 0.0005,threshold,0.0 ± 0.0,0.0021 ± 0.0007,transform,0.0014 ± 0.0006,0.0029 ± 0.0007,provabl,0.0 ± 0.0,0.0063 ± 0.0044,event,0.0016 ± 0.0005,0.0039 ± 0.0004
36,emb,0.0046 ± 0.0019,0.0069 ± 0.0008,discret,0.0012 ± 0.0009,0.0033 ± 0.0009,hash,0.0016 ± 0.0007,0.0031 ± 0.0007,exploit,0.0 ± 0.0,0.0062 ± 0.0044,agent,0.0009 ± 0.0004,0.0032 ± 0.0004
37,base,0.007 ± 0.0022,0.0093 ± 0.0009,regress,0.0043 ± 0.0016,0.0063 ± 0.0012,residu,0.0 ± 0.0,0.0015 ± 0.0005,power,0.0 ± 0.0,0.0062 ± 0.0044,recommend,0.0011 ± 0.0004,0.0033 ± 0.0004
38,noisi,0.0 ± 0.0,0.0022 ± 0.0006,answer,0.0017 ± 0.001,0.0038 ± 0.0009,prefer,0.0002 ± 0.0002,0.0016 ± 0.0005,stream,0.0 ± 0.0,0.0062 ± 0.0044,extract,0.0034 ± 0.0006,0.0056 ± 0.0005
39,privaci,0.0 ± 0.0,0.0022 ± 0.0006,addit,0.0 ± 0.0,0.002 ± 0.0008,predict,0.0116 ± 0.0014,0.013 ± 0.0011,mixtur,0.0 ± 0.0,0.0062 ± 0.0062,theori,0.0023 ± 0.0006,0.0045 ± 0.0005
40,commun,0.0 ± 0.0,0.0022 ± 0.0005,sourc,0.0 ± 0.0,0.002 ± 0.0007,approxim,0.0053 ± 0.0011,0.0067 ± 0.0009,prognosi,0.0 ± 0.0,0.0062 ± 0.0043,sentiment,0.0021 ± 0.0005,0.0042 ± 0.0004
41,multiclass,0.0012 ± 0.0012,0.0034 ± 0.0007,compar,0.0006 ± 0.0006,0.0026 ± 0.0008,pixel,0.0 ± 0.0,0.0014 ± 0.0006,kmean,0.0 ± 0.0,0.0061 ± 0.0043,social,0.002 ± 0.0005,0.004 ± 0.0004
42,infer,0.0046 ± 0.0019,0.0068 ± 0.0008,cluster,0.0123 ± 0.0028,0.0142 ± 0.0017,dialogu,0.0009 ± 0.0005,0.0022 ± 0.0006,task,0.0 ± 0.0,0.0061 ± 0.0044,tool,0.0003 ± 0.0002,0.0024 ± 0.0003
43,video,0.0009 ± 0.0009,0.003 ± 0.0006,domain,0.0017 ± 0.001,0.0036 ± 0.0009,ordin,0.0 ± 0.0,0.0014 ± 0.0005,neg,0.0 ± 0.0,0.006 ± 0.0043,cognit,0.0004 ± 0.0002,0.0025 ± 0.0004
44,sen,0.0 ± 0.0,0.0021 ± 0.0005,interpret,0.0 ± 0.0,0.0019 ± 0.0007,seri,0.002 ± 0.0006,0.0034 ± 0.0007,nonconvex,0.0 ± 0.0,0.0059 ± 0.0042,engin,0.0 ± 0.0,0.0021 ± 0.0003
45,markov,0.001 ± 0.001,0.003 ± 0.0006,transfer,0.0023 ± 0.0014,0.0043 ± 0.001,belief,0.0013 ± 0.0005,0.0026 ± 0.0006,memori,0.0 ± 0.0,0.0059 ± 0.0059,program,0.0053 ± 0.0008,0.0073 ± 0.0006
46,belief,0.0007 ± 0.0007,0.0028 ± 0.0006,context,0.0 ± 0.0,0.0019 ± 0.0006,express,0.0005 ± 0.0003,0.0018 ± 0.0005,guarant,0.0 ± 0.0,0.0059 ± 0.0042,concept,0.0003 ± 0.0002,0.0024 ± 0.0003
47,subspac,0.0 ± 0.0,0.0021 ± 0.0005,sequenc,0.0031 ± 0.0016,0.005 ± 0.0012,use,0.0164 ± 0.0014,0.0177 ± 0.0011,discoveri,0.0 ± 0.0,0.0059 ± 0.0041,search,0.0036 ± 0.0007,0.0057 ± 0.0005
48,output,0.0 ± 0.0,0.0021 ± 0.0006,label,0.002 ± 0.0011,0.0039 ± 0.0009,search,0.0036 ± 0.0008,0.0049 ± 0.0008,higherord,0.0 ± 0.0,0.0058 ± 0.0058,semant,0.0068 ± 0.0009,0.0089 ± 0.0006
49,spectral,0.0017 ± 0.0012,0.0037 ± 0.0007,semidefinit,0.0 ± 0.0,0.0019 ± 0.0007,pool,0.0003 ± 0.0003,0.0016 ± 0.0005,investig,0.0 ± 0.0,0.0058 ± 0.0041,databa,1e-04 ± 1e-04,0.0021 ± 0.0003
\end{filecontents*}
\csvautotabular{data/title_words_rej.csv}
		}
		\caption{Most important predictors for rejection in decreasing order of importance: title} \label{title_words_rej}
    \end{table*}
    \end{landscape}
    \clearpage
}

\afterpage{%
    \clearpage
    \thispagestyle{empty}
    \begin{landscape}
    \begin{table*}
        \centering 
        \resizebox{\columnwidth}{170pt}{
		\begin{filecontents*}{data/abstract_words_rej.csv}
,AI Stem,AI Score accepted,AI Score rejected,CL Stem,CL Score accepted,CL Score rejected,LG Stem,LG Score accepted,LG Score rejected,ICLR Stem,ICLR Score accepted,ICLR Score rejected,All Stem,All Score accepted,All Score rejected
0,adapt,0.0054 ± 0.0012,0.009 ± 0.0006,code,0.0017 ± 0.0006,0.0059 ± 0.0008,network,0.0263 ± 0.0013,0.0333 ± 0.0011,languag,0.007 ± 0.0022,0.0354 ± 0.0049,logic,0.0022 ± 0.0004,0.0091 ± 0.0005
1,robust,0.003 ± 0.0009,0.0065 ± 0.0005,pattern,0.0024 ± 0.0008,0.0061 ± 0.0008,classif,0.0137 ± 0.0009,0.0176 ± 0.0008,word,0.0105 ± 0.0036,0.0335 ± 0.0061,user,0.0046 ± 0.0006,0.0102 ± 0.0004
2,strategi,0.0031 ± 0.0009,0.0061 ± 0.0005,recognit,0.004 ± 0.001,0.0074 ± 0.0007,learn,0.0434 ± 0.0014,0.0471 ± 0.0011,relat,0.002 ± 0.0008,0.0155 ± 0.0035,detect,0.0041 ± 0.0004,0.0094 ± 0.0004
3,topic,0.0023 ± 0.0008,0.0053 ± 0.0006,structur,0.0122 ± 0.0018,0.0155 ± 0.001,accuraci,0.0089 ± 0.0006,0.0115 ± 0.0005,semant,0.0077 ± 0.003,0.0199 ± 0.004,speech,0.0022 ± 0.0004,0.0074 ± 0.0004
4,local,0.0048 ± 0.0012,0.0078 ± 0.0005,base,0.0142 ± 0.001,0.0174 ± 0.0007,train,0.0209 ± 0.001,0.0235 ± 0.0009,linguist,0.0003 ± 0.0003,0.012 ± 0.0028,concept,0.002 ± 0.0003,0.0069 ± 0.0004
5,approach,0.0175 ± 0.0015,0.0204 ± 0.0006,represent,0.0111 ± 0.0014,0.0142 ± 0.001,neural,0.0176 ± 0.001,0.0202 ± 0.0008,speech,0.0022 ± 0.001,0.0126 ± 0.0035,research,0.0039 ± 0.0003,0.0087 ± 0.0003
6,score,0.001 ± 0.0005,0.0039 ± 0.0004,stateoftheart,0.0058 ± 0.0007,0.0085 ± 0.0005,approxim,0.0106 ± 0.0008,0.013 ± 0.0008,annot,0.0008 ± 0.0006,0.0108 ± 0.0034,ontolog,0.0005 ± 0.0002,0.0053 ± 0.0004
7,defin,0.0024 ± 0.0006,0.005 ± 0.0004,semant,0.0054 ± 0.0011,0.008 ± 0.0008,rule,0.0033 ± 0.0006,0.0054 ± 0.0006,emot,0.0006 ± 0.0006,0.0103 ± 0.0045,paper,0.0129 ± 0.0003,0.0175 ± 0.0002
8,sgd,0.0006 ± 0.0004,0.0033 ± 0.0006,sampl,0.0128 ± 0.0016,0.0154 ± 0.0011,complex,0.0093 ± 0.0006,0.0114 ± 0.0006,model,0.036 ± 0.0042,0.0449 ± 0.0038,belief,0.0016 ± 0.0003,0.006 ± 0.0004
9,environ,0.0016 ± 0.0006,0.0042 ± 0.0004,continu,0.0023 ± 0.0005,0.005 ± 0.0006,attack,0.0011 ± 0.0003,0.0032 ± 0.0006,answer,0.0014 ± 0.0008,0.0097 ± 0.0036,plan,0.003 ± 0.0005,0.0074 ± 0.0005
10,agent,0.0029 ± 0.0013,0.0054 ± 0.0006,agent,0.0045 ± 0.0012,0.0071 ± 0.0009,addit,0.0054 ± 0.0005,0.0074 ± 0.0004,translat,0.0056 ± 0.0031,0.0134 ± 0.0037,fuzzi,0.0 ± 0.0,0.0043 ± 0.0004
11,reinforc,0.0027 ± 0.0009,0.0052 ± 0.0005,languag,0.0081 ± 0.0013,0.0107 ± 0.0009,cnn,0.0051 ± 0.0008,0.007 ± 0.0008,text,0.0086 ± 0.004,0.0164 ± 0.0031,robot,0.0018 ± 0.0003,0.0061 ± 0.0004
12,test,0.0059 ± 0.0011,0.0083 ± 0.0005,activ,0.0031 ± 0.0007,0.0057 ± 0.0007,local,0.0069 ± 0.0007,0.0087 ± 0.0006,question,0.0035 ± 0.0013,0.0108 ± 0.0034,intellig,0.0015 ± 0.0002,0.0055 ± 0.0003
13,set,0.0171 ± 0.0016,0.0196 ± 0.0006,latent,0.0047 ± 0.0011,0.0072 ± 0.0008,fuzzi,0.0008 ± 0.0003,0.0026 ± 0.0005,corpu,0.0002 ± 0.0002,0.0074 ± 0.0018,decis,0.0043 ± 0.0004,0.0083 ± 0.0004
14,plan,0.0011 ± 0.0007,0.0035 ± 0.0006,class,0.0094 ± 0.0013,0.0119 ± 0.001,layer,0.0089 ± 0.0009,0.0106 ± 0.0008,process,0.0087 ± 0.0017,0.0158 ± 0.0021,reason,0.004 ± 0.0004,0.0079 ± 0.0004
15,student,0.0 ± 0.0,0.0024 ± 0.0005,autoencod,0.0008 ± 0.0004,0.0032 ± 0.0006,regress,0.0059 ± 0.0008,0.0076 ± 0.0007,discours,0.0008 ± 0.0008,0.0077 ± 0.003,inform,0.0113 ± 0.0006,0.0152 ± 0.0004
16,embed,0.005 ± 0.0015,0.0073 ± 0.0007,topic,0.0038 ± 0.0011,0.0062 ± 0.0009,thi,0.0303 ± 0.0006,0.0321 ± 0.0005,web,0.0012 ± 0.0009,0.008 ± 0.0029,recognit,0.004 ± 0.0004,0.0078 ± 0.0003
17,region,0.0012 ± 0.0007,0.0035 ± 0.0004,current,0.003 ± 0.0006,0.0054 ± 0.0005,privaci,0.0016 ± 0.0005,0.0033 ± 0.0006,causal,0.0012 ± 0.0012,0.0079 ± 0.0044,discuss,0.0024 ± 0.0003,0.0061 ± 0.0003
18,context,0.0034 ± 0.0009,0.0057 ± 0.0004,dropout,0.0004 ± 0.0003,0.0028 ± 0.0008,power,0.0045 ± 0.0004,0.0061 ± 0.0004,import,0.0038 ± 0.0012,0.0105 ± 0.0018,describ,0.0042 ± 0.0003,0.0079 ± 0.0003
19,error,0.0074 ± 0.0013,0.0097 ± 0.0006,annot,0.0022 ± 0.0006,0.0046 ± 0.0007,gpu,0.0015 ± 0.0004,0.0032 ± 0.0005,english,0.0006 ± 0.0006,0.007 ± 0.002,web,0.0012 ± 0.0003,0.0049 ± 0.0003
20,dimens,0.0029 ± 0.0009,0.0051 ± 0.0005,causal,0.0018 ± 0.0009,0.0042 ± 0.001,input,0.0076 ± 0.0006,0.0092 ± 0.0006,separ,0.0021 ± 0.0013,0.0085 ± 0.0027,process,0.0091 ± 0.0005,0.0127 ± 0.0003
21,arm,0.0007 ± 0.0005,0.0029 ± 0.0005,encod,0.0028 ± 0.0007,0.0052 ± 0.0006,initi,0.0028 ± 0.0004,0.0044 ± 0.0004,sentenc,0.0047 ± 0.002,0.0111 ± 0.0025,use,0.0258 ± 0.0005,0.0292 ± 0.0003
22,learner,0.0018 ± 0.0009,0.0039 ± 0.0005,process,0.0101 ± 0.001,0.0124 ± 0.0007,theori,0.0034 ± 0.0004,0.005 ± 0.0004,task,0.0192 ± 0.0037,0.0255 ± 0.0027,knowledg,0.0082 ± 0.0006,0.0115 ± 0.0004
23,domain,0.0072 ± 0.0016,0.0093 ± 0.0007,bayesian,0.0059 ± 0.0011,0.0082 ± 0.0008,recurr,0.0056 ± 0.0006,0.0072 ± 0.0006,dictionari,0.0 ± 0.0,0.0063 ± 0.0034,softwar,0.0004 ± 1e-04,0.0037 ± 0.0003
24,word,0.0083 ± 0.002,0.0105 ± 0.0008,search,0.0055 ± 0.0011,0.0077 ± 0.0008,activ,0.0058 ± 0.0007,0.0074 ± 0.0006,address,0.0026 ± 0.0009,0.0088 ± 0.0016,rule,0.004 ± 0.0005,0.0072 ± 0.0004
25,achiev,0.0094 ± 0.0011,0.0114 ± 0.0004,label,0.0097 ± 0.0015,0.0119 ± 0.0012,dataset,0.0141 ± 0.0007,0.0156 ± 0.0007,extract,0.005 ± 0.0016,0.0112 ± 0.0024,techniqu,0.0082 ± 0.0005,0.0114 ± 0.0003
26,abil,0.0023 ± 0.0006,0.0044 ± 0.0003,text,0.0058 ± 0.0012,0.008 ± 0.0008,reduc,0.0057 ± 0.0005,0.0072 ± 0.0004,inform,0.0113 ± 0.002,0.0174 ± 0.0023,search,0.0057 ± 0.0006,0.0088 ± 0.0004
27,pca,0.0007 ± 0.0005,0.0027 ± 0.0004,speech,0.0018 ± 0.0006,0.004 ± 0.0007,binari,0.0037 ± 0.0006,0.0052 ± 0.0005,polit,0.0006 ± 0.0006,0.0065 ± 0.004,theori,0.0035 ± 0.0003,0.0065 ± 0.0003
28,problem,0.0249 ± 0.002,0.0269 ± 0.0007,memori,0.0031 ± 0.0008,0.0052 ± 0.0007,architectur,0.0083 ± 0.0008,0.0098 ± 0.0006,effect,0.0067 ± 0.0014,0.0126 ± 0.0018,classifi,0.006 ± 0.0005,0.009 ± 0.0004
29,rnn,0.0026 ± 0.0013,0.0046 ± 0.0006,regress,0.0056 ± 0.0013,0.0078 ± 0.0009,loss,0.0084 ± 0.001,0.0098 ± 0.0008,patient,0.0007 ± 0.0007,0.0065 ± 0.0034,semant,0.0095 ± 0.0007,0.0125 ± 0.0005
30,behavior,0.0023 ± 0.0009,0.0043 ± 0.0004,local,0.0062 ± 0.0011,0.0084 ± 0.0007,dimension,0.0043 ± 0.0006,0.0057 ± 0.0005,phrase,0.0014 ± 0.001,0.0072 ± 0.0026,review,0.0013 ± 0.0003,0.0041 ± 0.0003
31,covari,0.0005 ± 0.0003,0.0025 ± 0.0004,variat,0.0033 ± 0.0008,0.0054 ± 0.0007,convnet,0.0006 ± 0.0003,0.0019 ± 0.0005,use,0.0291 ± 0.0023,0.0347 ± 0.0024,event,0.0031 ± 0.0006,0.006 ± 0.0004
32,allow,0.0059 ± 0.0011,0.0078 ± 0.0004,knowledg,0.0066 ± 0.0011,0.0086 ± 0.0008,make,0.0067 ± 0.0004,0.0081 ± 0.0004,identifi,0.0045 ± 0.0016,0.0101 ± 0.0024,tool,0.0027 ± 0.0003,0.0053 ± 0.0003
33,program,0.0037 ± 0.001,0.0057 ± 0.0006,gene,0.0002 ± 0.0002,0.0023 ± 0.0007,adversari,0.0034 ± 0.0007,0.0048 ± 0.0006,similar,0.0093 ± 0.0027,0.0148 ± 0.0036,extract,0.0058 ± 0.0005,0.0084 ± 0.0003
34,deriv,0.0051 ± 0.001,0.007 ± 0.0004,possibl,0.0047 ± 0.0007,0.0067 ± 0.0006,manifold,0.0019 ± 0.0004,0.0033 ± 0.0005,audio,0.0024 ± 0.0017,0.0078 ± 0.0032,databas,0.0015 ± 0.0003,0.0041 ± 0.0003
35,framework,0.0093 ± 0.0012,0.0111 ± 0.0005,method,0.0282 ± 0.0018,0.0302 ± 0.0011,convolut,0.0091 ± 0.0009,0.0105 ± 0.0007,human,0.0051 ± 0.0018,0.0105 ± 0.0025,articl,0.0014 ± 0.0003,0.0039 ± 0.0002
36,transfer,0.0022 ± 0.0011,0.0041 ± 0.0006,pars,0.0007 ± 0.0003,0.0027 ± 0.0007,lstm,0.0028 ± 0.0005,0.0041 ± 0.0006,lexicon,0.0 ± 0.0,0.0054 ± 0.0025,pattern,0.003 ± 0.0004,0.0056 ± 0.0003
37,dropout,0.0011 ± 0.0007,0.0029 ± 0.0006,gpu,0.0004 ± 0.0002,0.0024 ± 0.0005,appli,0.0088 ± 0.0005,0.0101 ± 0.0005,charact,0.0 ± 0.0,0.0053 ± 0.002,program,0.0064 ± 0.0007,0.009 ± 0.0005
38,particular,0.0037 ± 0.0008,0.0055 ± 0.0003,dialogu,0.0 ± 0.0,0.002 ± 0.0006,predict,0.0155 ± 0.0011,0.0169 ± 0.0008,tag,0.0005 ± 0.0005,0.0057 ± 0.0026,social,0.0025 ± 0.0003,0.0048 ± 0.0003
39,simpl,0.007 ± 0.0012,0.0088 ± 0.0004,risk,0.0019 ± 0.0007,0.0038 ± 0.0006,deep,0.0161 ± 0.0011,0.0174 ± 0.0008,summar,0.0017 ± 0.0009,0.0067 ± 0.0035,base,0.0144 ± 0.0005,0.0167 ± 0.0003
40,path,0.0011 ± 0.0006,0.0029 ± 0.0005,decod,0.0003 ± 0.0003,0.0022 ± 0.0005,recommend,0.004 ± 0.0008,0.0053 ± 0.0007,align,0.0017 ± 0.0009,0.0066 ± 0.0031,autom,0.001 ± 0.0002,0.0034 ± 0.0002
41,difficult,0.0019 ± 0.0006,0.0037 ± 0.0003,converg,0.0067 ± 0.0012,0.0086 ± 0.0008,cifar,0.0017 ± 0.0004,0.003 ± 0.0004,plan,0.0016 ± 0.001,0.0065 ± 0.003,manag,0.0004 ± 1e-04,0.0028 ± 0.0002
42,cnn,0.0042 ± 0.0014,0.006 ± 0.0007,dataset,0.012 ± 0.0012,0.0139 ± 0.0009,allow,0.0069 ± 0.0005,0.0082 ± 0.0004,sourc,0.0035 ± 0.0012,0.0083 ± 0.002,text,0.0084 ± 0.0006,0.0108 ± 0.0004
43,detect,0.0063 ± 0.0016,0.0081 ± 0.0006,perform,0.0168 ± 0.0012,0.0187 ± 0.0007,theoret,0.0077 ± 0.0006,0.009 ± 0.0005,summari,0.0 ± 0.0,0.0048 ± 0.0026,technolog,0.0005 ± 1e-04,0.0028 ± 0.0002
44,batch,0.0011 ± 0.0006,0.0028 ± 0.0004,lstm,0.001 ± 0.0005,0.0028 ± 0.0006,depth,0.0013 ± 0.0004,0.0025 ± 0.0004,account,0.0015 ± 0.0008,0.0062 ± 0.0024,arab,1e-04 ± 1e-04,0.0024 ± 0.0003
45,superior,0.0014 ± 0.0006,0.0032 ± 0.0003,news,0.0003 ± 0.0002,0.0021 ± 0.0005,nonlinear,0.0037 ± 0.0005,0.005 ± 0.0005,experi,0.0094 ± 0.0016,0.0142 ± 0.0017,automat,0.0045 ± 0.0004,0.0068 ± 0.0002
46,outlier,0.0 ± 0.0,0.0017 ± 0.0004,rule,0.0048 ± 0.0014,0.0067 ± 0.0009,polici,0.0063 ± 0.0008,0.0075 ± 0.0008,embed,0.0069 ± 0.0025,0.0115 ± 0.0035,present,0.0126 ± 0.0005,0.0149 ± 0.0003
47,encourag,0.0005 ± 0.0003,0.0022 ± 0.0003,sentenc,0.0033 ± 0.001,0.0051 ± 0.0007,problem,0.0255 ± 0.001,0.0267 ± 0.0008,automat,0.0025 ± 0.0009,0.0071 ± 0.0016,engin,0.002 ± 0.0003,0.0042 ± 0.0003
48,solver,0.0012 ± 0.0005,0.0029 ± 0.0004,reward,0.0029 ± 0.0009,0.0047 ± 0.0007,review,0.0021 ± 0.0005,0.0033 ± 0.0006,natur,0.0079 ± 0.0015,0.0124 ± 0.0018,aim,0.0023 ± 0.0002,0.0045 ± 0.0002
49,variat,0.0039 ± 0.0012,0.0056 ± 0.0005,scheme,0.0042 ± 0.0009,0.006 ± 0.0006,outcom,0.001 ± 0.0003,0.0022 ± 0.0003,role,0.0002 ± 0.0002,0.0048 ± 0.0013,measur,0.0064 ± 0.0005,0.0086 ± 0.0004
\end{filecontents*}
\csvautotabular{data/abstract_words_rej.csv}
		}
		\caption{Most important predictors for rejection in decreasing order of importance: abstract} \label{abstract_words_rej}
    \end{table*}
    \end{landscape}
    \clearpage
}

\afterpage{%
    \clearpage
    \thispagestyle{empty}
    \begin{landscape}
    \begin{table*}
        \centering 
        \resizebox{\columnwidth}{170pt}{
		\begin{filecontents*}{data/introduction_words_rej.csv}
,AI Stem,AI Score accepted,AI Score rejected,CL Stem,CL Score accepted,CL Score rejected,LG Stem,LG Score accepted,LG Score rejected,ICLR Stem,ICLR Score accepted,ICLR Score rejected,All Stem,All Score accepted,All Score rejected
0,polici,0.0052 ± 0.0015,0.0094 ± 0.0009,user,0.0087 ± 0.0016,0.0136 ± 0.0012,learn,0.0498 ± 0.0013,0.0529 ± 0.0012,languag,0.0121 ± 0.0029,0.0349 ± 0.0045,section,0.0179 ± 0.0006,0.0271 ± 0.0004
1,detect,0.0056 ± 0.001,0.0096 ± 0.0007,bandit,0.004 ± 0.0011,0.008 ± 0.001,algorithm,0.0387 ± 0.0013,0.0417 ± 0.0012,word,0.0146 ± 0.0046,0.0368 ± 0.006,logic,0.0031 ± 0.0004,0.0105 ± 0.0005
2,graph,0.0093 ± 0.0022,0.0129 ± 0.001,embed,0.0056 ± 0.0011,0.0094 ± 0.001,architectur,0.0087 ± 0.0006,0.0116 ± 0.0007,plan,0.0015 ± 0.0007,0.0139 ± 0.0041,ontolog,0.0005 ± 1e-04,0.0066 ± 0.0005
3,spars,0.0066 ± 0.0014,0.0102 ± 0.0008,word,0.0137 ± 0.002,0.0174 ± 0.0014,approxim,0.013 ± 0.0008,0.0158 ± 0.0008,logic,0.0006 ± 0.0004,0.0129 ± 0.0036,concept,0.0031 ± 0.0003,0.0091 ± 0.0004
4,state,0.0082 ± 0.0011,0.0112 ± 0.0006,log,0.0029 ± 0.0006,0.0063 ± 0.0007,network,0.0319 ± 0.0015,0.0347 ± 0.0013,relat,0.0082 ± 0.001,0.0202 ± 0.003,user,0.009 ± 0.0007,0.0144 ± 0.0006
5,matrix,0.0135 ± 0.002,0.0164 ± 0.001,regret,0.0091 ± 0.0019,0.0124 ± 0.0013,sampl,0.0188 ± 0.0011,0.0215 ± 0.001,semant,0.0093 ± 0.0026,0.0208 ± 0.0034,describ,0.0057 ± 0.0002,0.0104 ± 0.0002
6,cost,0.0065 ± 0.0015,0.0095 ± 0.0007,document,0.0073 ± 0.0013,0.0106 ± 0.0012,random,0.0092 ± 0.0006,0.0118 ± 0.0006,event,0.0028 ± 0.001,0.0135 ± 0.0042,detect,0.0051 ± 0.0004,0.0099 ± 0.0004
7,model,0.0433 ± 0.0027,0.0459 ± 0.0011,mixtur,0.0034 ± 0.0008,0.0064 ± 0.0009,comput,0.0187 ± 0.0006,0.0212 ± 0.0006,translat,0.005 ± 0.0014,0.0156 ± 0.0034,research,0.0068 ± 0.0003,0.0116 ± 0.0002
8,recommend,0.0031 ± 0.0008,0.0058 ± 0.0007,translat,0.0044 ± 0.001,0.0073 ± 0.0009,action,0.0087 ± 0.0008,0.0111 ± 0.0009,web,0.0011 ± 0.0005,0.0114 ± 0.0038,inform,0.0136 ± 0.0004,0.0183 ± 0.0003
9,event,0.0032 ± 0.001,0.0057 ± 0.0008,arm,0.0036 ± 0.0016,0.0062 ± 0.0012,train,0.0268 ± 0.001,0.0291 ± 0.001,user,0.009 ± 0.0019,0.019 ± 0.0037,knowledg,0.0093 ± 0.0005,0.0138 ± 0.0004
10,featur,0.0245 ± 0.0029,0.027 ± 0.0012,search,0.0072 ± 0.0011,0.0099 ± 0.0008,classifi,0.0126 ± 0.0009,0.0147 ± 0.0009,annot,0.0031 ± 0.001,0.0119 ± 0.0031,fuzzi,0.0 ± 0.0,0.0043 ± 0.0005
11,decis,0.0056 ± 0.0012,0.0079 ± 0.0006,reward,0.0073 ± 0.0016,0.0099 ± 0.0011,data,0.0341 ± 0.0011,0.0362 ± 0.001,represent,0.0111 ± 0.0023,0.0199 ± 0.0025,semant,0.0111 ± 0.0006,0.0152 ± 0.0005
12,entiti,0.0033 ± 0.0011,0.0056 ± 0.0007,text,0.007 ± 0.0011,0.0095 ± 0.0009,convolut,0.0086 ± 0.0007,0.0105 ± 0.0008,inform,0.0139 ± 0.0017,0.022 ± 0.0023,web,0.0017 ± 0.0002,0.0057 ± 0.0004
13,reinforc,0.0029 ± 0.0006,0.0052 ± 0.0005,structur,0.0132 ± 0.0011,0.0158 ± 0.0009,search,0.0075 ± 0.0007,0.0094 ± 0.0006,ontolog,0.0006 ± 0.0005,0.0086 ± 0.003,speech,0.0048 ± 0.0004,0.0088 ± 0.0005
14,anomali,0.0 ± 0.0,0.0022 ± 0.0005,arab,0.0 ± 0.0,0.0024 ± 0.0008,layer,0.0141 ± 0.0011,0.0158 ± 0.001,question,0.0077 ± 0.0018,0.0156 ± 0.0037,plan,0.004 ± 0.0006,0.008 ± 0.0005
15,design,0.006 ± 0.0007,0.0082 ± 0.0004,memori,0.0042 ± 0.0007,0.0065 ± 0.0008,stream,0.002 ± 0.0003,0.0037 ± 0.0005,phrase,0.0021 ± 0.0012,0.01 ± 0.0032,event,0.0046 ± 0.0006,0.0083 ± 0.0005
16,action,0.0079 ± 0.0016,0.0101 ± 0.0009,depend,0.0089 ± 0.0007,0.0112 ± 0.0007,polici,0.0083 ± 0.001,0.01 ± 0.0009,knowledg,0.0087 ± 0.002,0.016 ± 0.0024,extract,0.0065 ± 0.0004,0.0101 ± 0.0004
17,patient,0.0014 ± 0.0006,0.0036 ± 0.0006,column,0.0022 ± 0.0006,0.0045 ± 0.0007,rate,0.0099 ± 0.0008,0.0116 ± 0.0008,formula,0.0004 ± 0.0002,0.0076 ± 0.0022,rule,0.0056 ± 0.0006,0.0092 ± 0.0004
18,base,0.0127 ± 0.0008,0.0147 ± 0.0004,compon,0.0047 ± 0.0005,0.0071 ± 0.0006,regress,0.0073 ± 0.0006,0.009 ± 0.0007,answer,0.0058 ± 0.0025,0.0128 ± 0.0035,agent,0.0105 ± 0.0008,0.0139 ± 0.0007
19,reward,0.0067 ± 0.0018,0.0087 ± 0.0008,lexic,0.0005 ± 0.0002,0.0028 ± 0.0006,target,0.0079 ± 0.0007,0.0096 ± 0.0006,verb,0.0 ± 0.0,0.0068 ± 0.003,program,0.008 ± 0.0008,0.0114 ± 0.0005
20,algorithm,0.0404 ± 0.003,0.0424 ± 0.0011,annot,0.0036 ± 0.0009,0.0058 ± 0.0007,point,0.0109 ± 0.0007,0.0125 ± 0.0006,transit,0.0011 ± 0.0004,0.0079 ± 0.0025,present,0.0094 ± 0.0003,0.0127 ± 0.0002
21,node,0.0067 ± 0.0017,0.0087 ± 0.0007,rbm,0.0023 ± 0.0009,0.0044 ± 0.001,matrix,0.0151 ± 0.0011,0.0167 ± 0.001,discours,0.0016 ± 0.0011,0.0084 ± 0.0039,social,0.0033 ± 0.0003,0.0067 ± 0.0004
22,categori,0.0033 ± 0.0006,0.0052 ± 0.0005,languag,0.0111 ± 0.0015,0.0132 ± 0.001,person,0.0026 ± 0.0003,0.0041 ± 0.0004,reason,0.0028 ± 0.0005,0.0093 ± 0.0016,reason,0.0061 ± 0.0003,0.0095 ± 0.0003
23,dual,0.0018 ± 0.0007,0.0038 ± 0.0005,gaussian,0.0055 ± 0.0009,0.0076 ± 0.0008,denot,0.0048 ± 0.0004,0.0064 ± 0.0004,claus,0.0005 ± 0.0003,0.007 ± 0.0029,softwar,0.0008 ± 1e-04,0.0042 ± 0.0003
24,gaussian,0.0051 ± 0.0011,0.0071 ± 0.0006,constraint,0.0073 ± 0.001,0.0094 ± 0.0008,stochast,0.0095 ± 0.0007,0.0111 ± 0.0007,emot,0.0016 ± 0.001,0.0081 ± 0.0037,intellig,0.0032 ± 0.0002,0.0065 ± 0.0003
25,causal,0.0011 ± 0.0007,0.003 ± 0.0007,solut,0.0088 ± 0.0008,0.0108 ± 0.0007,svm,0.0057 ± 0.0007,0.0072 ± 0.0008,solver,0.0013 ± 0.0007,0.0076 ± 0.0032,discuss,0.0051 ± 0.0002,0.0083 ± 0.0002
26,similar,0.0096 ± 0.0012,0.0116 ± 0.0006,entiti,0.0038 ± 0.001,0.0059 ± 0.0009,class,0.0129 ± 0.0008,0.0144 ± 0.0007,corpu,0.0014 ± 0.0007,0.0075 ± 0.0014,process,0.0113 ± 0.0004,0.0145 ± 0.0003
27,game,0.0048 ± 0.0016,0.0068 ± 0.0008,local,0.0081 ± 0.0009,0.0101 ± 0.0008,cnn,0.0086 ± 0.001,0.0101 ± 0.001,dynam,0.0019 ± 0.0006,0.0079 ± 0.0016,recognit,0.0061 ± 0.0003,0.0092 ± 0.0003
28,risk,0.0024 ± 0.0005,0.0044 ± 0.0005,case,0.0116 ± 0.0007,0.0135 ± 0.0005,sketch,0.0005 ± 1e-04,0.002 ± 0.0005,syntact,0.0008 ± 0.0005,0.0068 ± 0.0018,human,0.0083 ± 0.0005,0.0114 ± 0.0004
29,point,0.0103 ± 0.0013,0.0122 ± 0.0006,onlin,0.0085 ± 0.0014,0.0104 ± 0.0009,sgd,0.0038 ± 0.0007,0.0053 ± 0.0007,text,0.012 ± 0.0034,0.0179 ± 0.0034,servic,0.0011 ± 0.0002,0.0042 ± 0.0003
30,dynam,0.0047 ± 0.0007,0.0066 ± 0.0005,hyperparamet,0.0018 ± 0.0004,0.0037 ± 0.0008,weight,0.0108 ± 0.0008,0.0123 ± 0.0007,align,0.0013 ± 0.0006,0.0071 ± 0.0028,databas,0.0019 ± 0.0002,0.0049 ± 0.0003
31,item,0.0051 ± 0.0018,0.007 ± 0.0009,attribut,0.0033 ± 0.0009,0.0052 ± 0.0008,nois,0.0066 ± 0.0006,0.008 ± 0.0007,describ,0.0067 ± 0.0008,0.0124 ± 0.0013,technolog,0.0012 ± 1e-04,0.0042 ± 0.0002
32,norm,0.0056 ± 0.0012,0.0075 ± 0.0008,pca,0.0024 ± 0.0007,0.0043 ± 0.0008,game,0.0058 ± 0.0009,0.0072 ± 0.0008,mean,0.0056 ± 0.0011,0.0112 ± 0.0018,search,0.0077 ± 0.0005,0.0107 ± 0.0004
33,practic,0.0061 ± 0.0006,0.008 ± 0.0003,matrix,0.0147 ± 0.0018,0.0166 ± 0.0013,theorem,0.0042 ± 0.0004,0.0056 ± 0.0005,tree,0.0052 ± 0.0019,0.0108 ± 0.0032,text,0.0116 ± 0.0007,0.0145 ± 0.0005
34,manifold,0.0026 ± 0.0009,0.0044 ± 0.0007,project,0.0047 ± 0.0008,0.0065 ± 0.0006,multiclass,0.0025 ± 0.0005,0.0039 ± 0.0005,tag,0.0017 ± 0.0011,0.0072 ± 0.0025,organ,0.0028 ± 0.0002,0.0057 ± 1e-04
35,recoveri,0.0009 ± 0.0004,0.0027 ± 0.0004,summar,0.0026 ± 0.0004,0.0045 ± 0.0006,adversari,0.0053 ± 0.0009,0.0066 ± 0.0008,pars,0.0017 ± 0.0009,0.0071 ± 0.0021,formula,0.0012 ± 0.0002,0.0041 ± 0.0003
36,attack,0.0014 ± 0.0007,0.0032 ± 0.0006,tag,0.002 ± 0.0006,0.0038 ± 0.0007,low,0.0044 ± 0.0003,0.0057 ± 0.0004,retriev,0.0011 ± 0.0006,0.0064 ± 0.0019,decis,0.0059 ± 0.0004,0.0088 ± 0.0004
37,onlin,0.0087 ± 0.0017,0.0105 ± 0.0008,rule,0.0059 ± 0.0009,0.0077 ± 0.0009,perturb,0.0022 ± 0.0004,0.0035 ± 0.0005,articl,0.0023 ± 0.0008,0.0074 ± 0.0019,theori,0.0043 ± 0.0002,0.0072 ± 0.0003
38,data,0.0331 ± 0.0025,0.0349 ± 0.0009,score,0.0036 ± 0.0007,0.0053 ± 0.0006,neural,0.0185 ± 0.0009,0.0198 ± 0.0008,lexic,0.0014 ± 0.0009,0.0065 ± 0.0021,pattern,0.0042 ± 0.0004,0.007 ± 0.0003
39,thi,0.0423 ± 0.0012,0.0441 ± 0.0005,kmean,0.0027 ± 0.001,0.0045 ± 0.0009,column,0.0027 ± 0.0004,0.004 ± 0.0005,intellig,0.0017 ± 0.0006,0.0068 ± 0.0013,robot,0.0034 ± 0.0004,0.0062 ± 0.0004
40,hypothesi,0.0024 ± 0.0006,0.0042 ± 0.0004,track,0.0016 ± 0.0004,0.0033 ± 0.0006,gradient,0.0127 ± 0.001,0.014 ± 0.0008,state,0.0097 ± 0.0016,0.0148 ± 0.0032,person,0.0023 ± 0.0002,0.0052 ± 0.0003
41,famili,0.0025 ± 0.0006,0.0042 ± 0.0003,extract,0.0063 ± 0.001,0.008 ± 0.0006,attack,0.0027 ± 0.0006,0.004 ± 0.0007,descript,0.0018 ± 0.0006,0.0069 ± 0.0018,solver,0.0022 ± 0.0003,0.005 ± 0.0004
42,real,0.0037 ± 0.0005,0.0054 ± 0.0003,english,0.0009 ± 0.0003,0.0026 ± 0.0004,relu,0.0012 ± 0.0003,0.0024 ± 0.0005,pomdp,0.0003 ± 0.0003,0.0054 ± 0.0026,base,0.0127 ± 0.0003,0.0155 ± 0.0002
43,modul,0.0006 ± 0.0002,0.0022 ± 0.0003,mean,0.0065 ± 0.0006,0.0082 ± 0.0005,seri,0.0044 ± 0.0006,0.0056 ± 0.0007,databas,0.0017 ± 0.0008,0.0066 ± 0.0026,techniqu,0.0084 ± 0.0003,0.0111 ± 0.0003
44,known,0.0067 ± 0.0007,0.0083 ± 0.0003,cnn,0.0049 ± 0.0013,0.0066 ± 0.001,profil,0.0009 ± 0.0002,0.0021 ± 0.0004,arab,0.0 ± 0.0,0.0049 ± 0.0027,articl,0.0022 ± 0.0003,0.0048 ± 0.0003
45,infer,0.0102 ± 0.0018,0.0118 ± 0.0007,player,0.0026 ± 0.0008,0.0042 ± 0.0009,student,0.002 ± 0.0006,0.0032 ± 0.0006,evalu,0.0061 ± 0.001,0.0109 ± 0.0012,document,0.0078 ± 0.0006,0.0105 ± 0.0005
46,email,0.0016 ± 0.0005,0.0032 ± 0.0004,grid,0.0013 ± 0.0005,0.0029 ± 0.0006,reservoir,1e-04 ± 1e-04,0.0013 ± 0.0005,sentenc,0.0061 ± 0.0021,0.011 ± 0.0023,belief,0.0025 ± 0.0003,0.0051 ± 0.0004
47,elast,1e-04 ± 1e-04,0.0017 ± 0.0005,match,0.0048 ± 0.0008,0.0065 ± 0.0007,acceler,0.002 ± 0.0002,0.0032 ± 0.0004,document,0.0085 ± 0.0028,0.0133 ± 0.0031,environ,0.005 ± 0.0004,0.0076 ± 0.0004
48,dimension,0.0065 ± 0.001,0.0081 ± 0.0006,email,0.0019 ± 0.0004,0.0036 ± 0.0006,size,0.0078 ± 0.0005,0.009 ± 0.0004,theori,0.0038 ± 0.0009,0.0086 ± 0.0018,sat,0.0005 ± 1e-04,0.003 ± 0.0003
49,control,0.0053 ± 0.001,0.0069 ± 0.0006,descript,0.0022 ± 0.0004,0.0038 ± 0.0005,techniqu,0.0104 ± 0.0005,0.0116 ± 0.0005,domain,0.0097 ± 0.0034,0.0144 ± 0.0031,conclus,0.0012 ± 1e-04,0.0037 ± 1e-04
\end{filecontents*}
\csvautotabular{data/introduction_words_rej.csv}
		}
		\caption{Most important predictors for rejection in decreasing order of importance: introduction} \label{introduction_words_rej}
    \end{table*}
    \end{landscape}
    \clearpage
}

\end{document}